\documentclass[prd,preprint,superscriptaddress,preprintnumbers,eqsecnum,showpacs,nofootinbib,nobibnotes]{revtex4-1}
\usepackage{amsfonts,amsmath,amssymb,bm}
\allowdisplaybreaks[1] 
\usepackage[table]{xcolor}
\usepackage{graphicx} 
\usepackage{mathtools}
\usepackage{boondox-cal}
\usepackage{slashed}

\usepackage{xcolor}

\usepackage[utf8]{inputenc}

\newcommand{\be}{\begin{equation}}
\newcommand{\bea}{\begin{eqnarray}}
\newcommand{\ba}{\begin{align}}
\newcommand{\ee}{\end{equation}}
\newcommand{\eea}{\end{eqnarray}}
\newcommand{\ea}{\end{align}}

\definecolor{zero2}{rgb}{0.88,0.88,.88}

\def\1eq#1{Eq.~(\ref{#1})}

\def\2eqs#1#2{Eqs.~(\ref{#1}) and~(\ref{#2})}
\def\3eqs#1#2#3{Eqs.~(\ref{#1}),~(\ref{#2}) and~(\ref{#3})}
\def\4eqs#1#2#3#4{Eqs.~(\ref{#1}),~(\ref{#2}),~(\ref{#3}) and~(\ref{#4})}

\def\G{\Gamma}

\def\hphi0{{\hat\phi}_0}

\def\d{\!\mathrm{d}^4x\,}

\def\ln{{\mathcal{l}_h}}
\def\jn{\mathcal{j}}
\def\lS{{\mathcal{l}_h}}
\def\lV{{\mathcal{l}_{a'}}}

\def\Mp2{{\mu}^2}

\def\bpsi{\overline{\psi}}
\def\Ds{\slashed{D}}

\makeatletter
\def\user@resume{resume}
\def\user@intermezzo{intermezzo}
\newcounter{previousequation}
\newcounter{lastsubequation}
\newcounter{savedparentequation}
\def\CT@@do@color{%
	\global\let\CT@do@color\relax
		\@tempdima\wd\z@
		\advance\@tempdima\@tempdimb
		\advance\@tempdima\@tempdimc
		\advance\@tempdimb\tabcolsep
		\advance\@tempdimc\tabcolsep
		\advance\@tempdima1.5\tabcolsep
	\kern-1.5\@tempdimb
	\leaders\vrule
	\hskip\@tempdima\@plus  1fill
	\kern-1.5\@tempdimc
	\hskip-\wd\z@ \@plus -1fill }
\makeatother

\usepackage[english]{babel}

\def\G{\Gamma}
\def\lh{{\cal{l}_h}}
\def\la{{\cal{l}}_a}

\begin{document}

\title{
Gauge-invariant Slavnov-Taylor Decomposition\\
for Trilinear Vertices
}

\date{May 31st, 2025}

\author{A. Quadri}
\email{andrea.quadri@mi.infn.it}
\affiliation{INFN, Sezione di Milano, via Celoria 16, I-20133 Milano, Italy}

\begin{abstract}
\noindent
We continue the analysis of the gauge-invariant decomposition
of amplitudes in spontaneously broken massive gauge theories by providing a characterization
of separately gauge-invariant subsectors for  amplitudes involving trilinear 
interaction vertices for an Abelian
theory with chiral fermions.

\noindent
We show that the use of Frohlich-Morchio-Strocchi gauge-invariant dynamical (i.e. propagating inside loops) fields
yields a very powerful handle on the cancellations among unphysical degrees of freedom (the longitudinal
mode of the massive gauge field, the Goldstone scalar and the ghosts).

\noindent
The resulting cancellations are encoded into separately Slavnov-Taylor invariant
sectors for 1-PI amplitudes.
The construction works to all orders in perturbation theory.

This decomposition suggests a novel strategy for the determination of finite  counter-terms required to restore the Slavnov-Taylor identities in chiral theories in the absence of an invariant regularization scheme.

\end{abstract}


\maketitle

\section{Introduction}

Gauge-invariant fields of the Frohlich-Morchio-Strocchi (FMS) type~\cite{Frohlich:1980gj,Frohlich:1981yi} 
provide a consistent and gauge-invariant description
of spontaneously broken gauge theories~\cite{Maas:2017wzi}. They are related to the
original fields via an invertible field redefinition~\cite{Boeykens:2024bll}. This ensures the physical equivalence with the standard formulation 
in terms of elementary non-gauge-invariant fields, since the physical observables, including the S-matrix elements, do not depend on the specific
field parameterization chosen, provided that the asymptotic states are left unaffected. This is a general result of Quantum Field Theory known as  the Equivalence Theorem~\cite{Kamefuchi:1961sb,Bergere:1975tr,Blasi:1998ph,Ferrari:2002kz,Cohen:2024fak}. 
Physical equivalence can also be proven directly by cohomological arguments~\cite{Binosi:2022ycu}.

Gauge-invariant fields exhibit several desirable properties.
They are associated with positive spectral density~\cite{Dudal:2019pyg,Dudal:2020uwb},
as should be expected for a physical description of those excitations.
This is in contrast with the ordinary scalar field used to describe the Higgs 
mode, whose K\"all\'en-Lehmann representation is not gauge-invariant
and exhibits a non-positive spectral density.

Unfortunately power-counting renormalizability is in general not manifest when using gauge-invariant fields~\cite{Boeykens:2024bll}, due to their composite nature.
However it turns out that power-counting renormalizability can be retained provided one uses appropriate Lagrange multipliers embedded into the so-called BRST doublets~\cite{Barnich:2000zw,Quadri:2002nh}
in order to enforce the field change of variables, as discussed in~\cite{Quadri:2024aqo,Binosi:2022ycu,Binosi:2019olm}.

Moreover, gauge-invariant fields allow to control in an effective way the 
physical content of the Slavnov-Taylor (ST)~\cite{Slavnov:1972fg,Taylor:1971ff,Weinberg:1996kr}  identities, a crucial set of functional relations
for the vertex functional $\G$ generating the one-particle irreducible (1-PI) amplitudes of the model that  ensures the cancellation of unphysical intermediate states associated with the quartet mechanism~\cite{Becchi:1974xu,Kugo:1977zq,Curci:1976yb}.

In fact each  1-PI amplitude
can be expanded according to the number $\ln$ of
internal Higgs and the number $\lV$ of vector mesons gauge-invariant fields. 
It turns out that separate ST identities hold for the different sectors $(\ln,\lV)$, i.e. physical unitarity cancellations occur independently in each sector.

This natural yet technically  non-trivial property has been proven in Ref.~\cite{Quadri:2024aqo} within the Algebraic Renormalization approach~\cite{Piguet:1995er,Grassi:1999tp}. 
Explicit examples were presented for
ST identities involving the two-point 1-PI amplitudes in the Abelian Higgs-Kibble model with chiral fermions.

In the present paper we take one step further and extend the diagrammatic analysis of the FMS gauge-invariant formalism to ST identities involving three-point 1-PI amplitudes.

In particular we check the fulfillment of the sector-by-sector ST identities for the UV divergent parts of the fermionic amplitudes in an Abelian chiral model.

The main motivation is that for chiral theories no symmetric
regularization  scheme is known to preserve
the ST identities at the regularized level, due to the presence of the $\gamma_5$ matrix, and so one 
has to work out finite symmetry-restoring counter-terms order by order in the loop expansion~\cite{Belusca-Maito:2023wah}.

One may hope that the FMS gauge-invariant formalism can provide some guidance to identify a general strategy in order to get those finite counter-terms. 

As a preliminary step toward this goal, we check in the present paper the fulfillment of the ST identities for the UV divergent parts of the 1-PI amplitudes and describe in detail the underlying
cancellation mechanism. 

As is well-known, at one loop order UV divergences have to fulfill the ST identities, the ambiguities in the regularization procedure only affecting finite one-loop counter-terms.
Thus the UV divergent parts of the amplitudes provide
a very clean playground to understand the sector
decomposition holding true in the FMS formalism. 

First of all let us consider amplitudes without external
fermion legs.
For these amplitudes
it is clear that fermion loops always contribute to the $(0,0)$-sectors, while
$(\lS,\lV)$-sectors, where either $\lS$ or $\lV$ is greater than zero,
are unaffected by fermionic contributions.
So for those sectors one can safely use
naive Dimensional Regularization with an anti-commuting $\gamma_5$ matrix in order to ensure 
the fulfillment of the ST identities already at the regularized level.

On the other hand, one-loop amplitudes involving external fermion legs contain one fermionic chain. Therefore one in general gets contributions from those chains
to all $(\lS,\lV)$-sectors by the insertions on the chain of scalar, ghost and gauge propagators.

In this case several new aspects arise.  
Cancellations among unphysical states in these latter
sectors can be analytically traced and display an interesting structure, as will be explicitly shown.

Moreover, the relevant set of one-loop ST identities involve both
three-point and two-point 1-PI functions. 
However, their projection on the sectors
$(\lS,\lV)$ yields a set of algebraic relations where the two-point functions only appear in sectors with $\lS + \lV \leq 2$.
This is because at one loop there are obviously
at most two internal lines in a two-point 1-PI amplitude,
so these amplitudes cannot simply contribute
to sectors with $\lS + \lV = 3$ (at one loop order three point  functions have at most three internal lines).

This seemingly obvious fact has deep implications in the problem of restoration
of the ST identities broken by intermediate regularization and is tied
to the locality properties of the theory.

In fact finite counter-terms, required to restore the ST identities, can also be decomposed
according to the $(\lS,\lV)$-grading. Therefore finite counter-terms, modifying the two-point
1-PI amplitudes, will only affect the relevant ST identities projection to which
two-point amplitudes actually contribute. 
In the sector $\lS + \lV = 3$
one has to work out
finite counter-terms affecting the
three point functions directly.

We leave explicit results for the full set of finite symmetry-restoring counter-terms, derived by exploiting such a decomposition, to a separate publication.
The present paper is devoted to
describe the sector decomposition structure of the UV divergences.

The paper is organized as follows.
In Sect.~\ref{sec:model} we review the FMS gauge-invariant formalism in its BRST version and describe the 
ST identity decomposition on separately invariant sectors.
In Sect.~\ref{sec:recover} we make connection with the standard formalism and establish the dictionary between the FMS and the conventional amplitudes.
In Sect.~\ref{sec:2pt} we consider two-point ST identities.
Sect.~\ref{sec:3pts} contains the detailed analysis of selected fermionic three-point ST identities and their substructure, while Sect.~\ref{sec:sti.bos}
presents the analysis for ST identities involving 3-point bosonic amplitudes.
Finally conclusions are presented in Sect.~\ref{sec:concl}.
the appendices contain the tree-level vertex functional and the propagators
(Appendix~\ref{app:tl}), the functional identities of the theory (Appendix~\ref{app:fi}) and the explicit
results for the UV divergent part of the relevant
amplitudes involved (Appendix~\ref{app:uv}).

\section{The model}\label{sec:model}

We consider an Abelian Higgs-Kibble model 
with chiral fermions in the gauge-invariant formalism of~\cite{Quadri:2024aqo}.

The Frohlich-Morchio-Strocchi (FMS) gauge-invariant
fields are denoted by $h$ and $a_\mu$:
\begin{align}
& h \sim \frac{1}{v} \Big ( \phi^\dagger \phi - \frac{v^2}{2} \Big ) \, , \nonumber \\
& a_\mu \sim \frac{i}{ev^2} \Big [ 2 \phi^\dagger D_\mu \phi  - \partial_\mu ( \phi^\dagger \phi ) \Big ]
\end{align}
where $\sim$ denotes on-shell equivalence when the
equations of motions for the
Lagrange multipliers $X$ and $X_\mu$, enforcing the
relevant constraints on the fields $h, a_\mu$,
are imposed~\cite{Quadri:2024aqo}.

In the above equation
$\phi = \frac{1}{\sqrt{2}}( \sigma + v + i \chi)$ is the
Higgs complex field, $\sigma$ describes the physical scalar
mode and $\chi$ the Goldstone field. $v$ is the
vacuum expectation value (v.e.v.). 
Moreover 
$$ D_\mu = \partial_\mu \phi - i e A_\mu \phi$$
is the covariant derivative, $e$ being the coupling constant
and $A_\mu$ the Abelian gauge field.
$h,a_\mu$ are on-shell equivalent to the FMS gauge-invariant fields originally proposed
in~\cite{Frohlich:1980gj,Frohlich:1981yi}.

 At variance with the approach discussed in~\cite{Boeykens:2024bll}, power-counting
renormalizability remains manifest. 
The technical reason is to be ascribed to the presence
of suitable derivative operators multiplying the 
Lagrange multipliers once the latter are embedded into BRST doublets~\cite{Barnich:2000zw,Quadri:2002nh}. 
In perturbation theory
they do not spoil the equivalence to the original theory,
as can be seen both formally in the path-integral
formalism by integrating out the Lagrange multipliers
and in a mathematically rigorous approach
by BRST techniques~\cite{Binosi:2019olm,Quadri:2024aqo,Piguet:1995er}.

In cohomological language,
this is to say that the Lagrange multipliers only enter via a BRST-exact term~\cite{Quadri:2002nh}.

It turns out that the only dim.5 and dim.6 interaction terms 
in the classical vertex functional $\G^{(0)}$ in
Eq.(\ref{cl.vertex.functional}) are those proportional
to the Lagrange multipliers $X$ and $X_\mu$.
The propagators $\Delta_{X_\mu X_\nu}$
and $\Delta_{X_\mu A_\nu}$ are however vanishing, so that 
power-counting violating interaction
vertices involving $X_\mu$ are harmless.
Moreover the  potentially dangerous interaction term
$X \square \phi^\dagger \phi$ still does not violate
power-counting renormalizabilty, due to the fact that the propagators
$\Delta_{XX}$ and $\Delta_{X\sigma}$ fall off as $p^{-4}$,
as one can see from Eq.(\ref{scalar.symm.props}).

Thus the model offers a  power-counting renormalizable version
of the FMS construction.

The 1-PI vertex functional $\G$ obeys the following Slavnov-Taylor (ST) identity:
\begin{align}
    {\cal S} (\G) =
    \int \d \, \Big [ & \partial_\mu \omega 
    \frac{\delta \G}{\delta A_\mu} + 
    \frac{\delta \G}{\delta \sigma^*}
    \frac{\delta \G}{\delta \sigma} +
    \frac{\delta \G}{\delta \chi^*}
    \frac{\delta \G}{\delta \chi} \nonumber \\
    & +
    \frac{\delta \G}{\delta \bar \eta}
    \frac{\delta \G}{\delta \psi} +
    \frac{\delta \G}{\delta \eta}
    \frac{\delta \G}{\delta \bar \psi} 
    + b \frac{\delta \G}{\delta \bar \omega}
     \Big ] = 0 \, .
     \label{sti}
\end{align}
This functional identity holds as a consequence of the classical
BRST symmetry of the gauge-fixed classical action
\begin{align}
    & s A_\mu = \partial_\mu \omega \, , \qquad 
    s \sigma = - e \omega \chi \, , \qquad
    s \chi = e \omega (v + \sigma) \, , \qquad
    s \phi = i e \omega \phi \, , \nonumber \\
    & s \psi = -i \frac{e}{2} \gamma_5 \psi \omega \, ,
    \qquad s \bpsi = i \frac{e}{2} \omega \bpsi \gamma_5 \, , \qquad s \bar \omega = b \, , \qquad s b =0 \, ,
    \qquad s \omega = 0 \, .
    \label{gauge.brst}
\end{align}
$\omega$ is the ghost field, $\bar \omega$ the antighost.
$b$ is the Nakanishi-Lautrup field while $\psi, \bar \psi$
are chiral fermions. 

The BRST transformations of $\sigma, \chi, \psi, \bar \psi$,
being non-linear in the quantized fields, have to be defined
by coupling them in the tree-level vertex functional $\G^{(0)}$ to the antifields $\sigma^*, \chi^*, \bar \eta, \eta$~\cite{Gomis:1994he,Weinberg:1996kr}.
The complete classical vertex
functional
$\G^{(0)}$, including the antifield-dependent terms and external sources required to define the functional equations of motion for the Lagrange multipliers, is given in Eq.(\ref{cl.vertex.functional}).

$\G$ is the generating functional of the 1-PI amplitudes of the theory, namely 
\begin{align}
    \G = \sum_{n \geq 0} \G^{(n)} = \sum_{n \geq 0}  \sum_{\underline{r}} 
    \int \prod_{i} d^4 x_{r_i} \,
    c_{\underline{r}} \G^{(n)}_{\Phi_{r_1}(x_1) \dots \Phi_{r_m}(x_m)} ~\Phi_{r_1}(x_1) \dots \Phi_{r_m}(x_m) \, ,
\end{align}
where $\Phi$ stands for a field or an external source,
the sum is over the loop index $n$ and all $r$-tuples $\underline{r}$ of length $r=1,2,\dots$
and $c_{\underline{r}}$ is a combinatorial factor (for instance
in the case of three fields $\Phi_{r_1}=\Phi_{r_2}=\Phi_{r_3}$ of the same type, $c_{\underline{r}}=\frac{1}{3!}$).
$\G^{(n)}$ is the 1-PI generating functional of
1-PI amplitudes at order $n$ in the loop expansion.

One can further expand the 1-PI Green's functions according to the 
number of the gauge-invariant fields $h$ and the combination 
$a'_\mu$ given by Eq.(\ref{landau.vec.fred}), i.e.
\begin{align}
\G^{(n)}_{\Phi_{r_1}(x_1) \dots \Phi_{r_m}(x_m)} = \sum_{\lS,\lV} 
\G^{(n;\lS,\lV)}_{\Phi_{r_1}(x_1) \dots \Phi_{r_m}(x_m)} \, .
\end{align}
In the above equation we have denoted by
$\G^{(1)}_{\Phi_1 \dots \Phi_m}$ the functional derivative
\begin{eqnarray}
\left . 
\G^{(1)}_{\Phi_1 \dots \Phi_m} \equiv 
\frac{\delta^m \G^{(1)}}{\delta \Phi_1(x_1) \dots \delta \Phi_m(x_m)}
\right |_{\Phi=0}
\end{eqnarray}
and have suppressed for notation simplicity the space-time dependence of the fields and external sources $\Phi_i$.

$\G^{(n;\lS,\lV)}_{\Phi_{r_1}(x_1) \dots \Phi_{r_m}(x_m)}$ is the sum of 
$n$-loop diagrams contributing
to the 1-PI amplitude with external legs $\Phi_{r_1}(x_1), \dots, \Phi_{r_m}(x_m)$ 
and containing $\lS$ internal $h$-propagators and $\lV$ internal $a'_\mu$-propagators.
$\G^{(n;0,0)}_{\Phi_{r_1}(x_1) \dots \Phi_{r_m}(x_m)}$ is the sum of 
$n$-loop diagrams contributing to the 1-PI amplitude without internal
$h$- and $a'_\mu$-propagators.

One could in fact also choose not to use the combination $a'_\mu$ in Eq.(\ref{landau.vec.fred})  but use instead $a_\mu$ in Eq.(\ref{landau.diag}) and again obtain a decomposition of amplitudes
according to $\la$. We make the choice of $a'_\mu$ since Feynman rules in the diagonal basis are simplified with this choice.

The crucial result of~\cite{Quadri:2024aqo} is that 
the ST identities in Eq.(\ref{sti}) do not actually hold
only for the full 1-PI amplitudes $\G^{(n)}_{\Phi_{r_1}(x_1) \dots \Phi_{r_m}(x_m)}$, but separately for each $(\lS,\lV)$-sector, namely
\begin{align}
    \int \d \, \Bigg [ & \partial_\mu \omega 
    \frac{\delta \G^{(n;\ln,\lV)}}{\delta A_\mu} 
    + b \frac{\delta \G^{(n;\ln,\lV)}}{\delta \bar \omega} 
    \nonumber \\
    & + \frac{\delta \G^{(n;\ln,\lV)}}{\delta \sigma^*}
      \frac{\delta \G^{(0)}}{\delta \sigma}
    + \frac{\delta \G^{(0)}}{\delta \sigma^*}
      \frac{\delta \G^{(n;\ln,\lV)}}{\delta \sigma}
     + \frac{\delta \G^{(n;\ln,\lV)}}{\delta \chi^*}
      \frac{\delta \G^{(0)}}{\delta \chi}
    + \frac{\delta \G^{(0)}}{\delta \chi^*}
      \frac{\delta \G^{(n;\ln,\lV)}}{\delta \chi}  
      \nonumber \\
    &  + \frac{\delta \G^{(n;\ln,\lV)}}{\delta \bar \eta}
      \frac{\delta \G^{(0)}}{\delta \psi}
    + \frac{\delta \G^{(0)}}{\delta \bar \eta}
      \frac{\delta \G^{(n;\ln,\lV)}}{\delta \psi}  
     + \frac{\delta \G^{(n;\ln,\lV)}}{\delta  \eta}
      \frac{\delta \G^{(0)}}{\delta \bar \psi}
    + \frac{\delta \G^{(0)}}{\delta  \eta}
      \frac{\delta \G^{(n;\ln,\lV)}}{\delta \bar \psi}  
    \nonumber \\
    & + \sum_{k=1}^{n-1} \sum_{\jn = 0}^\ln 
    \sum_{\jn'=0}^{\lV} 
    \Big ( 
    \frac{\delta \G^{(k;\jn,\jn')}}{\delta \sigma^*}
    \frac{\delta \G^{(n-k;\ln - \jn,\lV- \jn')}}{\delta \sigma} + \frac{\delta \G^{(k;\jn,\jn')}}{\delta \chi^*}
    \frac{\delta \G^{(n-k;\ln - \jn,\lV- \jn')}}{\delta \chi}
    \nonumber \\
    & \qquad \qquad \qquad  +
    \frac{\delta \G^{(k;\jn,\jn')}}{\delta \bar \eta}
    \frac{\delta \G^{(n-k;\ln - \jn,\lV- \jn')}}{\delta \psi} +
    \frac{\delta \G^{(k;\jn,\jn')}}{\delta \eta}
    \frac{\delta \G^{(n-k;\ln - \jn,\lV- \jn')}}{\delta \bar \psi} \Big ) 
     \Bigg ] = 0 \, .
     \label{sti.decomp}
\end{align}
In the present paper we will specialize to the one loop order,
so that there are no contributions from the bilinear terms
in the last two lines of Eq.(\ref{sti.decomp}).

Moreover we will consider the Landau gauge $\xi=0$ and 
set the additional mass parameter~$m$ in Eq.(\ref{cl.vertex.functional}) equal to zero. There is no loss in generality in making such a choice.
$m$ is an additional parameter allowed by the symmetries of the theory
and the power-counting, yet it is unphysical and will disappear
once one recovers the amplitudes in the standard formalism,
as has been extensively checked in~\cite{Binosi:2019nwz}.

Thus one can conveniently set $m=0$ for the purpose of studying
the sector-by-sector ST identities fulfillment in the simplest possible setting.

In the Landau gauge ghosts are free, so there are no one loop corrections
to the BRST transformations of the fields of the theory, i.e.
\begin{align}
\G^{(1)}_{\sigma^* \dots } =
\G^{(1)}_{\chi^* \dots } = 
\G^{(1)}_{\eta \dots } = \G^{(1)}_{\bar \eta \dots } = 0 \, , 
\end{align}
where the dots stand for any number of internal fields and external sources of the model. 

Thus the tower of $(\lS,\lV)$-ST identities  simplifies to
\begin{align}
   \int \d \, \Bigg [ & \partial_\mu \omega 
    \frac{\delta \G^{(1;\ln,\lV)}}{\delta A_\mu} 
    + b \frac{\delta \G^{(1;\ln,\lV)}}{\delta \bar \omega} 
    \nonumber \\
    & 
    - e \omega \chi
      \frac{\delta \G^{(1;\ln,\lV)}}{\delta \sigma}
    + e \omega ( v + \sigma)
      \frac{\delta \G^{(1;\ln,\lV)}}{\delta \chi}  
    - i \frac{e}{2} \gamma_5 \psi \omega
      \frac{\delta \G^{(1;\ln,\lV)}}{\delta \psi}  
    + i \frac{e}{2} \omega \bar \psi \gamma_5 
      \frac{\delta \G^{(1;\ln,\lV)}}{\delta \bar \psi} 
      \Bigg ] = 0 \, .
\label{sti.decomp.1loop}
\end{align}

\section{Recovering the amplitudes in the conventional formalism}\label{sec:recover}

In order to recover the one-loop 1-PI amplitudes in the conventional formalism
we need to eliminate the auxiliary fields $X,h$ and $X_\mu, a_\mu$.
The procedure works as follows.
By Eqs.(\ref{app.jth.X.eoms}) and (\ref{app.jth.h.eoms}) we see that the $n$-loop vertex functional $\G^{(n)}$, $n \geq 1$ only depends
on the combinations
\begin{align}
  \widetilde{{\bar c}^*} = \bar c^* + (\square + m^2) X \, , \qquad
  \widetilde{{\bar c^*}_\mu} = {\bar c^*}_\mu +  \Sigma^{\mu\nu}_{(\xi)} X_\nu \, .
\end{align}
On the other hand, by going on-shell with $h, a_\mu$ at tree-level,
according to Eqs.(\ref{app.tree-level.h.eoms}) and (\ref{app.tree-level.anu.eoms}) we get respectively:
\begin{align}
    & \bar c^* + (\square + m^2) X = (M^2 - m^2) h = \frac{1}{v} (M^2 - m^2) \Big ( \phi^\dagger \phi - \frac{v^2}{2} \Big ) \, , \\
    & \bar c^*_\mu + \Sigma^{\mu\nu}_{(\xi)} X_\nu = 0 \, ,
    \label{oneloop.sols}
\end{align}
where in the first equation we have used the tree-level on-shell condition
\begin{align}
    \frac{\delta \G^{(0)}}{\delta X} = (\square + m^2) \Big [ h - \frac{1}{v} \Big ( \phi^\dagger \phi - \frac{v^2}{2} \Big ) \Big ] = 0\, .
    \label{X.tree-level.eom}
\end{align}
The latter equation has the solution 
\begin{align}
    h = \frac{1}{v} \Big ( \phi^\dagger \phi - \frac{v^2}{2} \Big ) \, 
\end{align}
that has been used in the first of Eqs.(\ref{oneloop.sols}).
The most general solution to Eq.(\ref{X.tree-level.eom}) is given by
\begin{align}
h = \frac{1}{v} \Big ( \phi^\dagger \phi - \frac{v^2}{2} \Big ) + \eta \, ,
\end{align}
$\eta$ being a Klein-Gordon massive field. However at the perturbative level one can
safely neglect the field $\eta$, as has been discussed at length in~\cite{Binosi:2019olm}.

We eventually conclude that at one-loop order going on-shell 
with the extended pairs of fields $X,h$ and $X_\mu, a_\nu$ in order to
make contact with amplitudes in the standard formalism amounts at 
i) disregarding all one-loop amplitudes involving $\bar c^*_\mu$, as a consequence of the condition in the second of Eqs.(\ref{oneloop.sols});
ii) replacing the source $\bar c^*$ with the combination
\begin{align}
\bar c^* \rightarrow \frac{1}{v} (M^2 - m^2) \Big ( \phi^\dagger \phi - \frac{v^2}{2} \Big ) \, .
\end{align}

\section{ST identities for two-point amplitudes}\label{sec:2pt}

We review here the analysis of the ST identities for two-point amplitudes.
Unlike in the computations presented in~\cite{Quadri:2024aqo}, where use 
has been made of internal off-diagonal propagators in the symmetric basis,
we recover here the same results by using the mass eigenstate propagators.

The reason is that mass eigenstate propagators are more suited to 
the automation of one- and higher loops computations via tools like FeynArts/FormCalc~\cite{Hahn:2000kx,Hahn:2000jm}.
Of course both bases are equivalent.
In order to switch from the mass eigenstate basis
to the symmetric one, Eqs.(\ref{scalar.symm2mass}) 
and (\ref{landau.vec.fred}) are used. One sees 
from these equations that
the symmetric Green functions involving only
$\sigma$, $A_\mu$ and $\chi$ coincide with those 
with external legs $\sigma'$, $A'''_\mu$ and $\chi$.


In the present paper we will check explicitly the fulfillment
of the ST identities sector by sector for the UV divergent part
of the relevant 1-PI amplitudes. We will work in Dimensional Regularization with anticommuting $\gamma^5$.
While this regularization scheme is known to be inconsistent (for a recent review see ~\cite{Belusca-Maito:2023wah}), it gives the correct results
for UV divergences at one loop order. In a separate publication we will discuss the renormalization
of the model in the Breitenlohner-Maison-'t Hooft'-Veltman (BMHV) scheme and the construction
of sector-by-sector finite counterterms restoring the ST identities broken by the intermediate
regularization.

The ST identities to be considered for two-point 1-PI functions
are obtained by taking one derivative
w.r.t. the ghost $\omega$ and then w.r.t. $A_\mu$ and $\chi$ respectively
and then setting all fields and external sources to zero.
We obtain the two identities
\begin{align}
& -\partial^\mu \G^{(1;\ln,\lV)}_{A_\mu A_\nu} + e v \G^{(1;\ln,\lV)}_{\chi A_\nu} = 0 \, , \nonumber \\
& -\partial^\mu \G^{(1;\ln,\lV)}_{A_\mu \chi} - e \G^{(1;\ln,\lV)}_{\sigma} + e v \G^{(1;\ln,\lV)}_{\chi\chi} = 0 \, ,
\label{sti.two.point}
\end{align}
to be verified in each $(\ln,\lV)$-sector.
Since they involve 2-point amplitudes, $\ln+\lV\leq 2$.

As a preliminary remark, we notice that in order to study the contributions from diagrams involving $\sigma'$ and $X'$ it is
useful to move to the variables
\begin{align}
\sigma_+ = \sigma' + X' \, , \qquad
\sigma_- = \sigma'- X' \, .
\end{align}
From Eq.(\ref{scalar.prop.diag}) one sees that the diagonal propagators $\Delta_{\sigma_+\sigma_+}$ and $\Delta_{\sigma_-\sigma_-}$ 
are vanishing while the off-diagonal propagator reads
\begin{align}
\Delta_{\sigma_- \sigma_+} = \frac{2 i}{p^2} \, .
\end{align}
Therefore non-vanishing contributions from diagrams involving $\sigma', X'$
can only arise by interaction vertices involving both $\sigma_+$ and $\sigma_-$.
Now the field $\sigma = \sigma ' + X' + h$ obviously depends only on $\sigma_+$ and $h$. 
By direct inspection one therefore sees that in the tree-level vertex functional Eq.(\ref{cl.vertex.functional}) all interaction vertices only depends on $\sigma$ (and thus on $\sigma_+$), with the exception
of the vertex (we specialize here to the case of interest $m=0$)
\begin{align}
-\frac{1}{v} X \square (\phi^\dagger \phi) =
\frac{1}{v} (X' + h) \square (\phi^\dagger \phi) =
\frac{1}{v} \Big [ \frac{1}{2} ( \sigma_+ - \sigma_- ) + h \Big ]\square (\phi^\dagger \phi) \supset - \frac{1}{2v} \sigma_- \square (\phi^\dagger \phi) \, .
\label{except.vertex}
\end{align}
Therefore all diagrams involving $\sigma'$ and $X'$ cancel out among themselves with the exception of those involving the vertex in Eq.(\ref{except.vertex}).

We now discuss in turn each identity, labeled by the 
fields over which functional differentiation is taken.
\subsection{$\omega A_\mu$-STI}

There are four sectors to be considered, namely $(0,0)$, $(1,0)$, $(0,1)$ and $(1,1)$.

\subsubsection{Sector $(0,0)$}

The $(0,0)$-sector is spanned by diagrams with no internal $h$ and $a'_\mu$-lines. 

\begin{figure}[ht]
    \centering
    \includegraphics[width=0.5\linewidth]{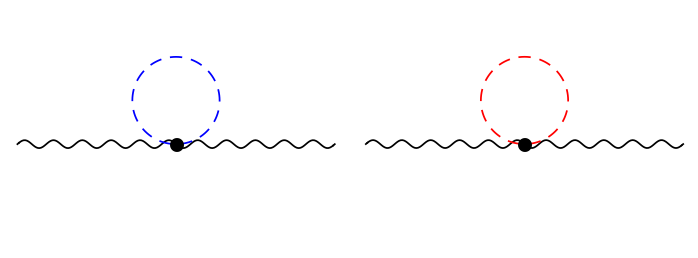}
    \caption{Diagrams contributing to 
    $\G^{(1;0,0)}_{A_\mu A_\nu}$ mutually compensating.
    The dashed red line represents the $\sigma'$-propagator,
    the dashed blue line the $X'$-propagator.
    }
    \label{fig.sti.AAcanc.00}
\end{figure}

Diagrams involving internal $\sigma'$
and $X'$-lines and no vertices of the type of Eq.~(\ref{except.vertex})
cancel against themselves and thus they can be safely dropped in the 
subsequent analysis.
For instance, diagrams in Fig.~\ref{fig.sti.AAcanc.00} cancel out since the 
interaction vertices are the same while the propagators of $\sigma'$
(dashed red line) and $X'$ (dashed blue line) have an opposite sign.

Moreover since we are in Landau gauge and the Goldstone fields are massless,
in Dimensional Regularization there are no contributions from Goldstone tadpoles.

\begin{figure}[ht]
    \centering
    \includegraphics[width=0.5\linewidth]{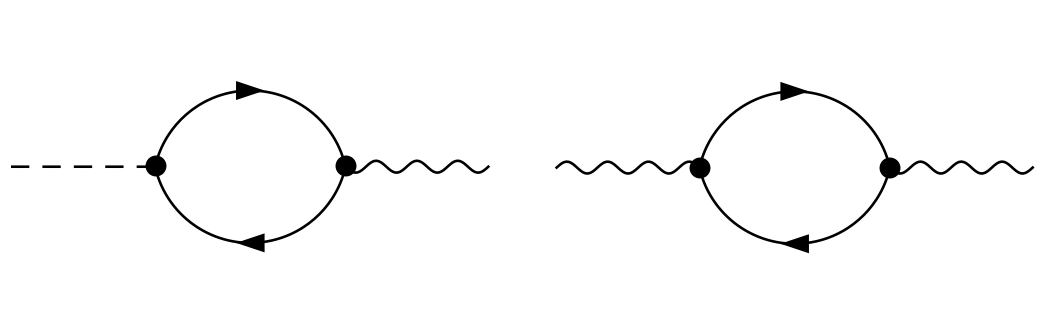}
    \caption{Non-vanishing amplitudes in $\G^{(1;0,0)}_{\chi A_\nu}$ and
    $\G^{(1;0,0)}_{A_\mu A_\nu}$.
    }
    \label{fig.AA-Achi.00}
\end{figure}

In Dimensional Regularization the only non-vanishing diagrams in $\G^{(1;0,0)}_{A_\mu A_\nu}$
and $\G^{(1;0,0)}_{\chi A_\nu}$ are the
fermionic bubbles depicted in Fig.~\ref{fig.AA-Achi.00}.

In momentum space the first of the ST identities in Eq.(\ref{sti.two.point}) in the $(0,0)$-sector can be written 
for the UV divergent parts of the 1-PI amplitudes, denoted by a bar, as
\begin{align}
    - i k^\mu \, \overline{\G}^{(1;0,0)}_{A_\mu(k)A_\nu(-k)} + ev \,\overline{\G}^{(1;0,0)}_{\chi(k) A_\nu(-k)} = 0 \, .
\end{align}
By direct inspection of Eqs.(\ref{G.Achi}) and (\ref{G.AA}) one can easily check
that the above relation is satisfied.

\subsection{Sectors $(1,0)$, $(0,1)$ and $(1,1)$}

The analysis of sectors $(1,0)$, $(0,1)$ and $(1,1)$ proceeds
in a similar way. No fermions are involved in these sectors.
By using 
(\ref{G.Achi}) and (\ref{G.AA})
one can easily verify that Eqs.(\ref{sti.two.point}) are fulfilled.

\subsection{$\omega \chi$-STI}

\begin{figure}
    \centering
    \includegraphics[width=0.5\linewidth]{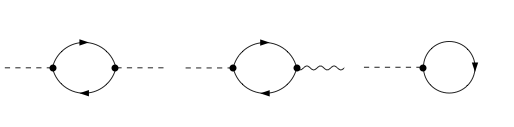}
    \caption{Fermion bubbles contributing to the $(0,0)$ sector
    of the $\omega\chi$-ST identity
    }
    \label{fig.sti.omega_chi}
\end{figure}

As in the case of $\omega A_\mu$-STI, there are four sectors to be considered, namely $(0,0)$, $(1,0)$, $(0,1)$ and $(1,1)$.
Fermion contributions appear only in the sector $(0,0)$ via
the diagrams depicted in Fig.~\ref{fig.sti.omega_chi}.
Again cancellations between diagrams involving $X'$, $\sigma'$ and no vertices in Eq.(\ref{except.vertex}) are at work.

By repeating the analysis and by direct inspection of 
Eqs.~(\ref{G.sigma}), (\ref{G.chichi}) and (\ref{G.Achi}), 
one can check that the ST identities
\begin{align}
-i k^\mu \overline{\G}^{(1;\ln,\lV)}_{A_\mu(k) \chi(-k)} - e \overline{\G}^{(1;\ln,\lV)}_{\sigma}
+ e v \overline{\G}^{(1;\ln,\lV)}_{\chi(k) \chi(-k)} = 0
\end{align}
are verified for each relevant $(\ln,\lV)$-sector.

We notice that, being a tadpole with just one internal line, $\overline{\G}^{(1;\ln,\lV)}_\sigma$
does not contribute to the sector
$(1,1)$.
This means that in this sector
the ST identity reduce to
\begin{align}
-i k^\mu \overline{\G}^{(1;1,1)}_{A_\mu(k) \chi(-k)} 
+ e v \overline{\G}^{(1;1,1)}_{\chi(k) \chi(-k)} = 0 \, ,
\end{align}
i.e. the highest order sector
only involves amplitudes with the same
greatest number of external legs. 
\medskip

A comment is in order here. The components
$(1,0)$ and $(1,1)$ of $\overline{\G}_{A_\mu \chi}$ 
and $\overline{\G}_{\chi\chi}$ respectively
contain
dimension 3 terms $\sim p^2 p_\mu$
and dimension 4 terms $\sim p^4$ 
that  violate power-counting renormalizability. 

However in the sum $\overline{\G}^{(1)}_{A_\mu \chi} = \sum_{\ln,\lV} 
\overline{\G}^{(1;\ln,\lV)}_{A_\mu \chi} $ 
and
$\overline{\G}^{(1)}_{\chi \chi} = \sum_{\ln,\lV} 
\overline{\G}^{(1;\ln,\lV)}_{\chi \chi} $ 
they cancel out against
themselves.
This is a consequence of the fact that power-counting renormalizability
is violated in each sector, yet the sum of all sectors must
obey bounds imposed by power-counting renormalizability, 
since the full theory is power-counting renormalizable.

\section{Slavnov-Taylor identities for fermionic amplitudes}\label{sec:3pts}

By taking a functional derivative of Eq.(\ref{sti.decomp.1loop}) w.r.t. $\omega$,
$\psi$ and $\bar \psi$ and then setting all fields and external sources to zero one obtains the set of identities
in momentum space
\begin{align}
- i k^\mu \G^{(1;\lh,\lV)}_{\psi(p) \bar \psi(q) A_\mu(k)} + i e \gamma^5 \G^{(1;\lh,\la)}_{\psi(p)\bar\psi(q)} + e v \G^{(1;\lh,\lV)}_{\psi(p) \bar \psi(q) \chi(k)} = 0 \,  .
\label{sti.omega_psi_barpsi}
\end{align}
At one loop order the relevant sectors to be considered are 
$(1,1)$, $(1,0)$, $(0,1)$ and $(0,0)$.
Let us consider each of them in detail.

\subsection{Sector $(1,1)$}

 This sector is spanned by all diagrams involving one internal $a'_\mu$-propagator and one internal $h$-propagator. There are no such diagrams contributing to the two-point function $\G^{(1)}_{\psi\bar\psi}$, so the ST identity reduces to
 \begin{align}
 - i k^\mu \G^{(1;1,1)}_{\psi(p) \bar \psi(q) A_\mu(k)} + e v \G^{(1;1,1)}_{\psi(p) \bar \psi(q) \chi(k)} = 0 \,  .
    \label{sti.omega_psi_barpsi.11}
 \end{align}
\begin{figure}
    \centering
    \includegraphics[width=0.5\linewidth]{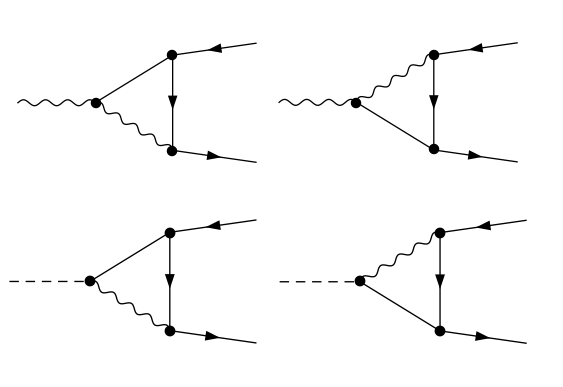}
    \caption{Diagrams contributing to the $(1,1)$-sector in amplitudes
    $\G^{(1;1,1)}_{\psi \bar \psi A_\mu}$ (first line) and 
    $\G^{(1;1,1)}_{\psi \bar \psi \chi}$ (second line). The internal wavy lines are $a'_\mu$-propagators, the solid lines $h$-propagators.
    }
    \label{fig.sti.omega_psi_barpsi.11}
\end{figure}
The diagrams to be considered are depicted in Fig.~\ref{fig.sti.omega_psi_barpsi.11}.
We observe first of all  that by Eq.~(\ref{landau.diag}) the symmetric function
$\G^{(1)}_{A_\mu \psi \bar \psi}$ is equal to the
amplitude in the mass eigenstate basis $\G^{(1)}_{A^{'''}_\mu \psi \bar \psi}$.

In order to see how the cancellations embodied in Eq.(\ref{sti.omega_psi_barpsi.11}) work diagrammatically, we notice that
the fermion chain is the same in both diagrams in each column in Fig~\ref{fig.sti.omega_psi_barpsi.11}.

Thus we can focus on the product of the trilinear vertices with the 
$a'_\mu$-propagator.
One finds from Eq.(\ref{cl.vertex.functional}), after going to the
mass eigenstate basis
\begin{align}
\G^{(0)}_{h(p) \chi(k) a'_\mu(-p-k) } = \frac{1}{ev^2}
\left . \Sigma^{\mu\nu}_{(0)}(-p -k) \right |_{M_A=0} (k-p)_\nu\, 
\label{fr.h_ap_chi}
\end{align}
and 
\begin{align}
\G^{(0)}_{h(p) A^{'''}_\nu(k) a'_\mu(-p-k) } = -\frac{2i}{v}
\left . \Sigma^{\mu\nu}_{(0)}(-p -k) \right |_{M_A=0} \, .
\label{fr.h_ap_A'''}
\end{align}

In addition one has to take into account the $ev$ prefactor in
the second term of the Slavnov-Taylor identity Eq.(\ref{sti.omega_psi_barpsi.11}).
So one finds ($p$ is the momentum of the $h$-propagator, incoming into the vertex)
\begin{align}
& ev \G^{(0)}_{h(p) \chi(k) a'_\mu(-p-k) } \Delta_{a'_\mu a'_\rho}  = 
\frac{1}{v} \left .  \Sigma^{\mu\nu}_{(0)}(-p -k) \right |_{M_A=0} (k-p)_\nu \Delta_{a'_\mu a'_\rho} \nonumber \\
& \qquad \qquad \qquad= 
\frac{1}{v} \left [ \Sigma^{\mu\nu}_{(0)}(-p -k) - M_A^2 g^{\mu\nu} \right ]\Delta_{a'_\mu a'_\rho}(k - p)
_\nu
\nonumber \\
& \qquad \qquad \qquad= \frac{i}{v} (k - p)_\rho - \frac{M_A^2}{v} \Delta_{a'_\nu a'_\rho}(k - p)_\nu \, ,
\label{sti.omega_psi_barpsi.contr.1}
\end{align}
where in the second line we have used the fact that 
$\Sigma^{\mu\nu}_{(0)} \Delta_{a'_\mu a'_\rho} = i g^{\nu}_{\rho}$.

Similarly one finds for the diagrams with an external 
gauge line contracted with the incoming momentum $k$
\begin{align}
-i k_\nu \Big ( -\frac{2i}{v}  \left . \Sigma^{\mu\nu}_{(0)} \right |_{M_A=0} \Big ) \Delta_{a'_\mu a'_\rho}
= -i \frac{2}{v} k_\rho + \frac{2}{v} M_A^2 k_\mu  \Delta_{a'_\mu a'_\rho} \, .
\label{sti.omega_psi_barpsi.contr.2}
\end{align}
By summing Eqs.(\ref{sti.omega_psi_barpsi.contr.1})
and (\ref{sti.omega_psi_barpsi.contr.2}) one gets
\begin{align}
-\frac{i}{v} (k + p)_\rho + \frac{M_A^2}{v} \Delta_{a'_\nu a'_\rho} ( k+p)_\nu = 0
\end{align}
since 
\begin{align}
\Delta_{a'_\nu a'_\rho} ( k+p)_\nu  = \frac{i}{M_A^2} (k+p)_\rho \, .
\end{align}
This shows that the ST identity Eq.(\ref{sti.omega_psi_barpsi.11}) is indeed satisfied.

Explicit values of the UV divergent parts of the relevant amplitudes
are reported in Appendix~\ref{app:uv}. It is interesting to see that also
these diagrams are indeed UV-divergent, as a consequence of the
derivative interactions in the classical action Eq~(\ref{cl.vertex.functional}).
For instance, the form factors UV coefficients $E_0^{(1;\ln,\lV)}$
reported in Eq.~(\ref{G.sigmaAA}) exhibit individually
a more severe degree of divergence than expected in a power-counting renormalizable theory (according to which $E_0$ should have UV degree at most $1$, in fact
zero by Lorentz-covariance).
$E_0^{(1;1,0)}$, $E_0^{(1;2,0)}$,
$E_0^{(1;2,1)}$ contain terms of order $\sim p^2$,
$E_0^{(1;1,1)}$ and $E_0^{(1;1,2)}$ terms of order $p^4$.

Nevertheless the sum of all contributions
$\sum_{\ln,\lV}~E_0^{(1;\ln,\lV)}$ must obey the power-counting renormalizable bound and thus must have UV dimension zero. One can check by direct inspection
of Eq.(\ref{G.sigmaAA}) that this is indeed the case.
This is a highly non-trivial check that must hold true as a consequence of the special manifest
power-counting renormalizability of the 
present formalism.

\subsection{Sectors $(0,0), (0,1)$ and $(1,0)$}

In these sectors also the two-point amplitude $\G^{(1)}_{\bar \psi \psi}$ contributes. The analysis can be carried out along
the same lines  for all sectors
$(0,0)$, $(0,1)$ and $(1,0)$.
Also here for general arguments one can aptly resort to the symmetric basis, while automated explicit computations for
each sector are carried out in the diagonal basis.

\begin{figure}[ht]
    \centering
    \includegraphics[width=0.5\linewidth]{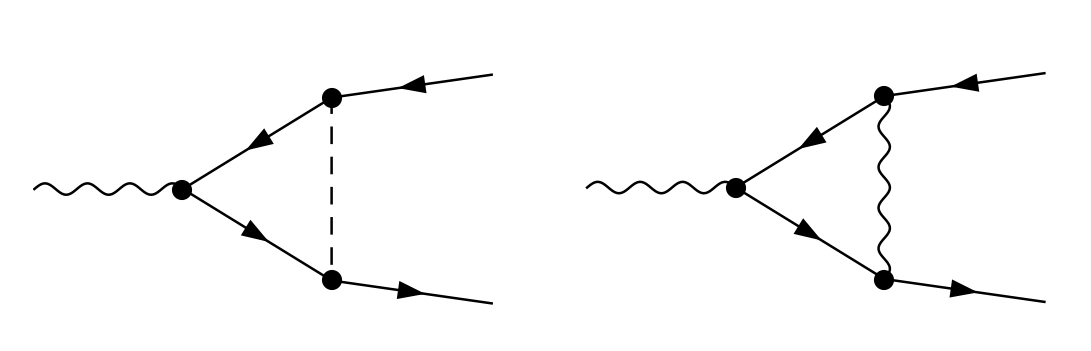}
    \caption{Diagrams with three fermion interaction vertices.
    The dashed line stands  for the $h, \sigma, \chi$-propagators,
    the wavy line denotes the $A_\mu$-propagator.
    }
    \label{fig.3ferm}
\end{figure}

Let us start from diagrams with three fermion interaction vertices, as depicted in Figure~\ref{fig.3ferm}.

They can be written schematically as
\begin{align}
    {\cal A}^{(\ln,\lV)}_{\chi(p) \psi(p_1) \bar \psi(p_2)}
    & =  i \int \frac{d^D q}{(2 \pi)^D} \, \G^{(0)}_{\psi \bar \psi \varphi} 
    \frac{\slashed{q} + m_e}{q^2 - m_e^2}
        \G^{(0)}_{\psi \bar \psi \chi} 
    \frac{\slashed{q}+\slashed{p} + m_e}{(q+p)^2 - m_e^2}
     \G^{(0)}_{\psi \bar \psi \varphi} \Delta^{(\ln,\lV)}_{\varphi\varphi}(q+p+p_1)  \, ,
     \nonumber \\
    {\cal A}^{(\ln,\lV)}_{A_\mu(p) \psi(p_1) \bar \psi(p_2)}
    & =  i \int \frac{d^D q}{(2 \pi)^D} \, \G^{(0)}_{\psi \bar \psi \varphi} 
    \frac{\slashed{q} + m_e}{q^2 - m_e^2}
        \G^{(0)}_{\psi \bar \psi A_\mu} 
    \frac{\slashed{q}+\slashed{p} + m_e}{(q+p)^2 + m_e^2}
     \G^{(0)}_{\psi \bar \psi \varphi} \Delta^{(\ln,\lV)}_{\varphi\varphi}(q+p+p_1)  \, ,  
     \label{amp.ferm.3}
\end{align}
where $\varphi$ can be either $h$, $\chi$, $A_\mu$ and $\sigma$. Notice that the fermions only interact with the scalar $\sigma$ and not with $X$, so there are no diagrams involving $\Delta_{X\sigma}$.

The combination
\begin{align}
-i k^\mu {\cal A}^{(\ln,\lV)}_{A_\mu \psi \bar \psi}
+ ev {\cal A}^{(\ln,\lV)}_{\chi \psi \bar \psi} 
\end{align}
can be simplified by using the tree-level ST identity
\begin{align}
   -i k^\mu \G^{(0)}_{A_\mu \psi \bar \psi}
+ ev \G^{(0)}_{\chi \psi \bar \psi}  = - i e \gamma_5 \G^{(0)}_{\psi \bar \psi} =  - i e \gamma_5 ( \slashed{p} - m_e )
\end{align}
on the relevant tree-level 1-PI Green's 
in the first two amplitudes
of Eq.(\ref{amp.ferm.3}).

By taking into account that 
\begin{align}
- i e \gamma_5  \G^{(0)}_{\psi \bar \psi} \frac{ (\slashed{q} + \slashed{p} ) + m_e }{(q+p)^2 - m_e^2}  = 
- i e  \gamma_5 \frac{(  \slashed{q} + \slashed{p}  - m_e ) 
( \slashed{q}+\slashed{p} + m_e  )}{(q+p)^2 - m_e^2}  
= - i e \gamma_5 
\end{align}
and momentum conservation $p+p_1+p_2=0$
one ends up with the identity
\begin{align}
-i k^\mu {\cal A}^{(\ln,\lV)}_{A_\mu \psi \bar \psi}
+ ev {\cal A}^{(\ln,\lV)}_{\chi \psi \bar \psi} + i e \gamma_5  {\cal A}^{(\ln,\lV)}_{\psi \bar \psi} = 0 \, 
\end{align}
where
\begin{align}
  {\cal A}^{(\ln,\lV)}_{\psi \bar \psi(p_1)} 
    =  i \int \frac{d^D q}{(2 \pi)^D} \, \G^{(0)}_{\psi \bar \psi \varphi}  
    \frac{\slashed{q} + m_e}{q^2 - m_e^2}
     \G^{(0)}_{\psi \bar \psi \varphi} \Delta^{(\ln,\lV)}_{\varphi\varphi}(q+p_1) \, , 
\end{align}

i.e. the ST identity holds true for this particular subset of diagrams in the sector $(\ln,\lV)$.
 
\begin{figure}[ht]
    \centering
    \includegraphics[width=0.5\linewidth]{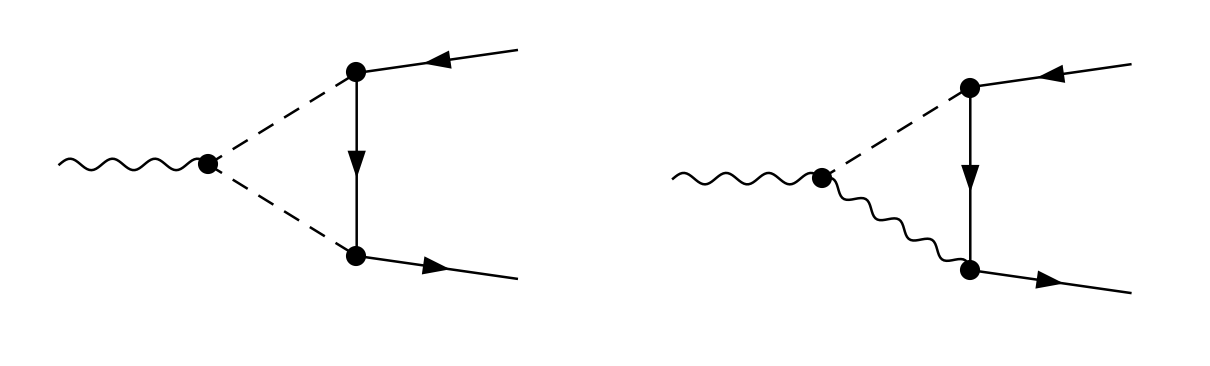}
    \caption{Diagrams with two fermion interaction vertices.
    In the left diagrams one line is always a Goldstone $\chi$, while the second one can stand for $\sigma, X, h$ propagators. The wavy line denotes the $A_\mu$-propagator.
    }
    \label{fig.2ferm}
\end{figure}

The analysis of diagrams with two
fermion interaction vertices displayed in Fig.~\ref{fig.2ferm}
follows the same lines as described
in the previous Subsection.

One can check explicitly the
fulfillment of the ST identity
for the UV divergent parts of the relevant
 amplitudes, sector by sector,
 by using the results 
 reported in Appendix~\ref{app:uv}.
 
\section{Slavnov-Taylor identities for bosonic amplitudes}\label{sec:sti.bos}

We list here the set of Slavnov-Taylor identities involving the trilinear
interaction vertices obtained by taking one derivative w.r.t. $\omega$, $\sigma$, $A_\nu$ and $\omega$, $\sigma$, $\chi$ respectively:
\begin{align}
    & - \partial^\mu \G^{(1)}_{\sigma A_\nu A_\mu} + ev \G^{(1)}_{\sigma A_\nu \chi} 
    + e \G^{(1)}_{A_\nu \chi} = 0 \, , \nonumber\\
    & -\partial^\mu \G^{(1)}_{\sigma \chi A_\mu} + ev \G^{(1)}_{\sigma \chi \chi} 
    + e \G^{(1)}_{\chi \chi} - e \G^{(1)}_{\sigma \sigma} = 0 \, .
    \label{sti.bosonic}
\end{align}
Each Slavnov-Taylor identity in Eqs.(\ref{sti.bosonic}) can be further decomposed
into sectors 
\begin{align}
    & - \partial^\mu \G^{(1;\ln,\lV)}_{\sigma A_\nu A_\mu} + ev \G^{(1;\ln,\lV)}_{\sigma A_\nu \chi} 
    + e \G^{(1;\ln,\lV)}_{A_\nu \chi} = 0 \, ,  \nonumber \\
    & -\partial^\mu \G^{(1;\ln,\lV)}_{\sigma \chi A_\mu} + ev \G^{(1;\ln,\lV)}_{\sigma \chi \chi} 
    + e \G^{(1;\ln,\lV)}_{\chi \chi} - e \G^{(1;\ln,\lV)}_{\sigma \sigma} = 0 \, .
    \label{sti.bosonic.sect}
\end{align}

The various contributions can be classified according to the value of the sum $\lS + \lV$.
Since the ST identities involve trilinear amplitudes, this sum can be at most $3$.
In this latter case the possible contributions are classified according to the sectors 
$(3,0)$, $(2,1)$, $(1,2)$, $(0,3)$.

One sees by the Feynman rules of the theory that there are no contributions
from the sectors $(3,0)$ and $(0,3)$. Moreover there are no contributions
from the two point amplitudes to the sectors where $\lS + \lV=3$ (because obviously the two point amplitudes at one loop
can have at most two internal lines).

When $\lS + \lV =2$, the relevant sectors are $(2,0)$, $(1,1)$ and $(0,2)$.
There are contributions to all the amplitudes, as can be seen from the
explicit computations of the UV divergent part of those amplitudes
reported in Appendix~\ref{app:uv}.

When $\lS + \lV=1$, the sectors to be considered are
$(1,0)$ and $(0,1)$.
Finally there is the  $(0,0)$-sector, containing
all amplitudes without any internal $h$ and $a'_\mu$-lines,
including the fermion bubbles.

In momentum space the first of Eqs.~(\ref{sti.bosonic.sect}) is given by
\begin{align}
    - i q^\mu \G^{(1;0,0)}_{\sigma(k) A_\nu(p) A_\mu(q)} 
    + e v \G^{(1;0,0)}_{\sigma(k) A_\nu(p) \chi(q)} +
    e \G^{(1;0,0)}_{A_\nu(p)\chi(q)} = 0 \, .
    \label{sti.b.2.mom}
\end{align}

Eqs.(\ref{sti.bosonic.sect}) read
\begin{align}
    - i q^\mu \G^{(1;0,0)}_{\sigma(k) \chi(p) A_\mu(q)} 
    + e v \G^{(1;0,0)}_{\sigma(k) \chi(p) \chi(q)} +
    e \G^{(1;0,0)}_{\chi(p)\chi(q)} - e \G^{(1)}_{\sigma(k) \sigma(q)} = 0 \, .
    \label{sti.b.2.mom.four}
\end{align}
Momentum conservation implies that $k+p+q=0$ (all momenta are considered
to be incoming) and that the support of the two-point functions
is restricted to the hyperplane $p+q=0$ for the two-point $\chi$ amplitude
and $A_\mu\chi$ amplitude
and to the hyperplane $k+q=0$ for the two point $\sigma$ amplitude.
We have explicitly verified all the relevant ST identities in each
$(\ln,\lV)$ sectors by direct computations. 

The results for the UV divergent terms
are reported in Eqs~(\ref{G.chichi}), (\ref{G.sigmasigma}), (\ref{G.sigmachichi}), 
(\ref{G.Achi}), 
(\ref{G.sigmaAA}) and
(\ref{G.Achisigma}).

As far as finite symmetry-restoring counter-terms are concerned, the above remarks imply that
for the highest value of the sum $\lS + \lV$
equals $3$,
only trilinear amplitudes must be corrected.
This fact is not manifest in the
full ST identity, see 
Eqs.(\ref{sti.bosonic}).

Similarly when it comes to sectors with $\lS+\lV=2$,
no tapdole contributions can enter.

By exploiting this fact one can arrange finite counter-terms in layers and have a guidance in order to evaluate them sector by sector. Detailed results are left to a separate publication.


\section{Conclusions}\label{sec:concl}

In this paper we have illustrated  the fulfillment of ST identities
sector by sector in the gauge-invariant variables by focussing on the UV divergent terms
of the relevant amplitudes at one loop order.
We have investigated the simplest non-trivial ST identities involving trilinear vertices
both for bosonic and for fermionic amplitudes. 

The explicit evaluation of the UV divergences of the theory shows that
power-counting renormalizability is lost within each sector, due to the presence
of higher derivative interactions, yet the model remains power-counting renormalizable
so that the sum of the different sector decomposition yields a 
manifestly power-counting renormalizable amplitude.

This is a very nice and peculiar property of the gauge invariant formalism proposed in~\cite{Quadri:2024aqo}, that does not hold in general for models implementing
gauge invariant FMS fields~\cite{Boeykens:2024bll}.

Since the ST identities hold true sector by sector, one can renormalize each sector
independently. In particular sectors without fermions do not suffer from the 
absence of a consistent invariant regularization scheme in chiral theories.
Notice that the notion of gauge-invariant sectors holds true to all orders
in perturbation theory and provides the appropriate mathematical framework
to discuss separately invariant subsets of diagrams. 

We have also shown that for ST identities only involving external bosonic legs (and possibly ghosts, in gauges different than Landau), the fermionic contributions are restricted
to the $(0,0)$-sector.
On the other hand, amplitudes with external fermion legs present contributions to all admissible sectors. 

Therefore a consistent regularization scheme, fulfilling the ST identities in a chiral gauge theory, can be defined by using naive dimensional regularization in all sectors
unaffected by the presence of fermions and by adopting e.g. the BMHV scheme in the remaining sectors plus the corrections due to finite symmetry-restoring counter-terms.

The explicit evaluation of such counter-terms will be presented in a separate publication, as well as
the extension of the formalism to non-Abelian gauge theories.

\section*{Acknowledgments}

Useful discussions with Dominik Stockinger and Thomas Hahn are gratefully acknowledged.

\appendix

\section{Tree-level effective action}\label{app:tl}

The full tree-level effective action is 
\begin{align}
    \G^{(0)} = \int \d  \Big \{ &
        - \frac{1}{4} F_{\mu\nu}^2 + (D^\mu \phi)^\dagger D_\mu \phi -
        \frac{m^2}{2v^2} \Big ( \phi^\dagger \phi - \frac{v^2}{2} \Big )^2 + i \bpsi \Ds \psi + \frac{G}{\sqrt{2}} \bpsi ( 1 - \gamma_5) \psi \phi + {\mbox{h.c.}} \nonumber \\
        & 
        -\frac{M^2-m^2}{2}h^2
        + \frac{\xi}{2} b^2 - b \Big ( \partial A + \xi M_A \chi \Big ) + \bar \omega \square \omega + \xi e^2 v \bar \omega \omega ( v + \sigma ) \nonumber \\
        & + X (\square + m^2) \Big [ h -  \frac{1}{v} \Big ( \phi^\dagger \phi - \frac{v^2}{2} \Big ) \Big ] - \bar c (\square + m^2) c \nonumber \\
        & - \bar c_\mu \Sigma_{(\xi)}^{\mu\nu} c_\nu +
        X_\mu \Sigma_{(\xi)}^{\mu\nu} 
        \Big [ a_\nu - \frac{i}{ev^2} \Big ( 2 \phi^\dagger  D_\nu \phi - \partial_\nu ( \phi^\dagger \phi ) \Big ) \Big ] \nonumber \\
        & - \sigma^* e \omega \chi + \chi^* e \omega ( v + \sigma) - \frac{ i e }{2} \bar \eta  \gamma_5 \psi \omega + \frac{ i e }{2} \omega \bar \psi \gamma_5 \eta \nonumber \\
        & + \bar c^* \Big [ h -  \frac{1}{v} \Big ( \phi^\dagger \phi - \frac{v^2}{2} \Big ) \Big ]
        + \bar c^*_\mu \Big [ a^\mu - \frac{i}{ev^2} \Big ( 2 \phi^\dagger  D^\mu \phi - \partial^\mu ( \phi^\dagger \phi ) \Big )\Big ]
    \Big \} \, .
    \label{cl.vertex.functional}
\end{align}
The operator $\Sigma_{(\xi)}^{\mu\nu}$ 
in Eq.(\ref{cl.vertex.functional}) is 
given by 
\begin{align}
\Sigma_{(\xi)}^{\mu\nu} \equiv \Big [ \square g^{\mu\nu} - \Big ( 1 - \frac{1}{\xi} \Big ) \partial^\mu \partial^\nu \Big ] + M_A^2 g^{\mu\nu} 
\end{align}
for $\xi \neq 0$ and by
\begin{align}
    \Sigma_{(0)}^{\mu\nu} \equiv (\square g^{\mu\nu} - \partial^\mu \partial^\nu) + M_A^2 g^{\mu\nu}
\end{align}
for the $\xi=0$ case.
In the present paper we will focus only on the Landau
gauge $\xi=0$.

Diagonalization of the quadratic part in the scalar sector is obtained by the following field redefinitions:
\begin{align}
    X = -X' - h \, , \qquad \sigma = \sigma' + X' + h \, .
    \label{scalar.fred}
\end{align}
The propagators in the mass eigenstates basis
are
\begin{align}
    \Delta_{\sigma'\sigma'} = - \Delta_{X'X'} = 
    \frac{i}{p^2 - m^2} \, , \qquad
    \Delta_{hh} = \frac{i}{p^2 - M^2} \, .
    \label{scalar.prop.diag}
\end{align}
One gets back the original symmetric basis via the field
transformations
\begin{align}
    X'= -X - h \, , \qquad \sigma'= \sigma + X \, .
    \label{scalar.symm2mass}
\end{align}
The  propagators in the symmetric basis are
\begin{align}
    \Delta_{XX} = \Delta_{X\sigma} = - \frac{i (M^2 - m^2)}{(p^2 - M^2) (p^2 - m^2)} \, , \qquad
    \Delta_{hh} = \Delta_{\sigma h} = \Delta_{\sigma \sigma} = - \Delta_{X h} = \frac{i}{p^2 - M^2} \, .
    \label{scalar.symm.props}
\end{align}

In a similar fashion diagonalization of the
vector-scalar quadratic part is achieved by the following
field redefinition~\cite{Quadri:2024aqo}
\begin{align}
& b = b' + M_A \chi \, , \qquad A_\mu = A''_\mu+X''_\mu+a'_\mu - \frac{1}{M_A^2}\partial_\mu b'\, , \nonumber \\
& X_\mu = X''_\mu + a'_\mu \, , \qquad
a_\mu = a'_\mu - \frac{1}{M_A^2}\partial_\mu b' - \frac{1}{M_A} \partial_\mu \chi \, .
\label{landau.diag}
\end{align}
The non-vanishing mass eigenstate propagators are given by
\begin{align}
    \Delta_{A''_\mu A''_\nu} = \Delta_{a'_\mu a'_\nu} = - \Delta_{X''_\mu X''_\nu} = \frac{i}{-p^2 + M_A^2} T_{\mu\nu} + \frac{i}{M_A^2}L_{\mu\nu} \, , 
    ~~
    \Delta_{b'b'} = -\frac{i M_A^2}{p^2} \, , ~~
    \Delta_{\chi\chi} = \frac{i}{p^2} \, .
    \label{landau.gauge.props}
\end{align}
We notice that the dependence on $b'$ of $A_\mu$
in Eq.(\ref{landau.diag}) is via the combination
$A'''_\mu = A''_\mu - \frac{1}{M_A^2}\partial_\mu b'$,
so one can avoid introducing interaction vertices
depending on $b'$ by using the field $A'''_\mu$ instead
of $A''_\mu$. Still propagators are diagonal and 
one has
\begin{align}
\Delta_{A'''_\mu A'''_\nu} = \frac{i}{-p^2 + M_A^2} T_{\mu\nu} \, ,
\end{align}
i.e. a purely transverse propagator, as expected.
The inverse field transformation is given by
\begin{align}
    & b' = b - M_A \chi \, , \qquad
       A'''_\mu = A_\mu - X_\mu \, , \nonumber\\ 
    & X''_\mu = X_\mu - a_\mu - \frac{1}{M_A^2} \partial_\mu b \, , \qquad 
    a'_\mu = a_\mu + \frac{1}{M_A^2} \partial_\mu b \, .
    \label{landau.vec.fred}
\end{align}

We notice by the last of Eqs.(\ref{landau.vec.fred}) that $a'_\mu$ is also gauge-invariant.

We adopt these mass eigenstate fields in the automation
of the calculation of 1-PI amplitudes by using FeynArts/FormCalc~\cite{Hahn:2000kx,Hahn:2000jm}.

\section{Functional Identities}\label{app:fi}

The vertex functional $\G$ is the generating functional of the 1-PI amplitudes
of the theory. It can be expanded according to the loop order $n$ as
\begin{align}
    \G = \sum_{j \geq 0} \G^{(j)} \, .
    \label{app.vert.funct}
\end{align}
$\G^{(0)}$ is the tree-level effective action (including the external sources
for composite operators) given in Eq.(\ref{cl.vertex.functional}).

\begin{itemize}
    \item The $X$- and $h$-equations of motion

    The tree-level effective action obeys the following identities:
    \begin{align}
        & \frac{\delta \G^{(0)}}{\delta X} = (\square + m^2) \frac{\delta \G^{(0)}}{\delta \bar c^*} \, , 
        \label{app.tree-level.X.eoms}
        \\
        & \frac{\delta \G^{(0)}}{\delta h} = (\square + m^2) X - ( M^2 - m^2) h + \bar c^* \, .
        \label{app.tree-level.h.eoms}
    \end{align}
    Since the r.h.s of the above equations is linear in the quantized fields,
    both identities hold true at the quantum level, i.e.
    \begin{align}
        & \frac{\delta \G^{(j)}}{\delta X} = (\square + m^2) \frac{\delta \G^{(j)}}{\delta \bar c^*} \, , \qquad j \geq 1 \, , 
        \label{app.jth.X.eoms} 
        \\
        & \frac{\delta \G^{(j)}}{\delta h} = 0 \, , \qquad j \geq 1 \, .
        \label{app.jth.h.eoms} 
    \end{align}
    The first equation implies that the whole dependence of the
    vertex functional on $X$ happens via the combination
    $\widetilde{{\bar c}^*} = \bar c^* + (\square + m^2) X$, while
    the dependence on the field~$h$ is confined at tree-level (in the symmetric basis we are using here).
    \item The $X_\mu$- and $a_\mu$-equations of motion
    $\G^{(0)}$ also obeys the following identities:
    \begin{align}
        & \frac{\delta \G^{(0)}}{\delta X_\mu} = \Sigma^{\mu\nu}_{(\xi)} \frac{\delta \G^{(0)}}{\delta \bar c^*_\nu} \, , 
        \label{app.tree-level.Xmu.eoms}
        \\
        & \frac{\delta \G^{(0)}}{\delta a_\mu} = \Sigma^{\mu\nu}_{(\xi)} X_\nu + \bar c^*_\mu \, . 
        \label{app.tree-level.anu.eoms}
    \end{align}
    By a similar argument they translate at the quantum level into
    \begin{align}
        & \frac{\delta \G^{(j)}}{\delta X_\mu} = \Sigma^{\mu\nu}_{(\xi)} \frac{\delta \G^{(j)}}{\delta \bar c^*_\nu} \, , \qquad j \geq 1 \, , 
        \label{app.jth.Xmu.eoms} \\
        & \frac{\delta \G^{(j)}}{\delta a_\mu} = 0 \, , \qquad j \geq 1 \, . 
        \label{app.jth.anu.eoms}
    \end{align}
    The last equation implies that the dependence on the field~$a_\mu$ is confined at tree-level (in the symmetric basis we are using here), while the first one
    entails that the dependence on $X_\mu$ only happens via the combination
    $\widetilde{{\bar c^*}_\mu} = {\bar c^*}_\mu +  \Sigma^{\mu\nu}_{(\xi)} X_\nu$, to all orders in the loop expansion.
\end{itemize}

\section{Sector decomposition of UV divergences}\label{app:uv}

We list here the sector decomposition of the UV divergent parts 
for the relevant 1-PI amplitudes. UV divergent parts are denoted by the same amplitude barred.

For fermionic amplitudes the chirality projectors are defined as
\begin{align}
    P_L = \frac{1}{2}( 1 - \gamma_5) \, , \qquad
    P_R = \frac{1}{2}( 1 + \gamma_5) \, , 
\end{align}
while $\sigma^{\mu\nu}=\frac{i}{2} [\gamma^\mu, \gamma^\nu]$.
All Dirac matrices are understood in $D=4$. 
Amplitudes are evalutaed in Dimensional Regularization with $\epsilon=4-D$. 

\subsubsection{The tadpole $\G^{(1)}_\sigma$}
\begin{alignat}{2}
\overline{\G}_\sigma^{(1;0,0)}
  &= -\,\frac{m_e^4}{2\pi^2\,v}\,\frac{1}{\epsilon}, 
&\quad
\overline{\G}_\sigma^{(1;1,0)}
  &= \frac{M^4}{8\pi^2\,v}\,\frac{1}{\epsilon}, 
\nonumber\\
\overline{\G}_\sigma^{(1;0,1)}
  &= 
  \frac{3\,e \,M_A^3}{8\pi^2}\,\frac{1}{\epsilon},
&\quad
\overline{\G}_\sigma^{(1;1,1)}
  &= 0 \, .
  \label{G.sigma}
\end{alignat}


%

\subsubsection{The two-point Goldstone amplitude $\G^{(1)}_{\chi\chi}$}

\begin{alignat}{2}
    \overline{\G}_{\chi\chi}^{(1;0,0)} &=
        \frac{m_e^2}{4\pi^2\,v^2}\,\frac{p^2-2m_e^2}{\epsilon}\, ,
    &\quad
    \overline{\G}_{\chi\chi}^{(1;1,0)}
    &= \frac{M_A^2 M^4  + 3 M^2 (M^2 + M_A^2)\,p^2 \;-\; M_A^2\,p^4}
        {8\,\pi^2\,e^2\,v^4}
        \;\frac{1}{\epsilon} \, , \nonumber \\
    \overline{\G}_{\chi\chi}^{(1;0,1)}
        &=\frac{3\,e^2\,M_A^2}{8\pi^2}\,\frac{1}{\epsilon} \, ,
    &\quad
    \overline{\G}_{\chi\chi}^{(1;1,1)}
        &= \frac{-3 M^4 -3 M_A^2 M^2 -3 M_A^4 + M_A^2\,p^2 }
        {8\,\pi^2\,e^2\,v^4}\,\frac{p^2}{\epsilon} \, .
\label{G.chichi}
\end{alignat}

\subsubsection{The two-point Higgs amplitude $\G^{(1)}_{\sigma\sigma}$}

\begin{alignat}{2}
\overline{\G}_{\sigma\sigma}^{(1;0,0)}
  &= -\,\frac{24\,m_e^4 \;-\;4\,m_e^2\,p^2 \;+\;3\,p^4}
       {16\,\pi^2\,v^2}\;\frac{1}{\epsilon} \, ,
       &\quad 
       \overline{\G}_{\sigma\sigma}^{(1;0,1)} &=
       -\,\frac{21\,M_A^4 + 9\,M_A^2\,p^2 - 2\,p^4}
       {8\,\pi^2\,v^2}\,\frac{1}{\epsilon} \, , \nonumber \\
\overline{\G}_{\sigma\sigma}^{(1;1,0)}
  &= -\,\frac{M^2\bigl(3M^2 + p^2\bigr)}{8\pi^2\,v^2}\,\frac{1}{\epsilon} \, ,
       & \quad 
       \overline{\G}_{\sigma\sigma}^{(1;1,1)} &= 0
\, , \nonumber \\
\overline{\G}_{\sigma\sigma}^{(1;2,0)}
  &=  \frac{M^2\,(6M^2 + p^2)}{8\pi^2\,v^2}\,\frac{1}{\epsilon} \, ,
       & \quad 
       \overline{\G}_{\sigma\sigma}^{(1;0,2)} &= \frac{60\,M_A^4 + 12\,M_A^2\,p^2 - p^4}
     {16\,\pi^2\,v^2}
\;\frac{1}{\epsilon} \, .
 \nonumber \\
      \label{G.sigmasigma}
\end{alignat}

\subsubsection{The two-point mixed Goldstone-gauge field amplitude $\G^{(1)}_{A_\mu \chi(p)}$}

\begin{alignat}{2}
\overline{\G}_{A_\mu \chi(p)}^{(1;0,0)}
  &=
\frac{i\,e\,m_e^2\,p_{\mu}}{4\pi^2\,v}\,\frac{1}{\epsilon}\, ,
&\quad
\overline{\G}_{A_\mu \chi(p)}^{(1;1,0)}
  &= i 
\frac{\, 3M^2(M^2 + M_A^2) - M_A^2\,p^2}{8\pi^2 \,e\,\,v^3}\,\frac{p_{\mu}}{\epsilon}
  \, , \nonumber \\
\overline{\G}_{A_\mu \chi(p)}^{(1;0,1)}
  &=0 \, ,
&\quad
\overline{\G}_{A_\mu \chi(p)}^{(1;1,1)}
  &= -i \frac{\,3\,(M^4 + M_A^2 M^2  + M_A^4) - M_A^2\,p^2}
      {8\pi^2 \,e\,v^3}\,\frac{p_\mu}{\epsilon}\, .
        \label{G.Achi}
\end{alignat}

\subsubsection{The two-point gauge field amplitude $\G^{(1)}_{A_\mu A_\nu}$}
\begin{alignat}{2}
& \overline{\G}_{A_\mu A_\nu}^{(1;0,0)}
  =
\frac{e^2}{24\pi^2}\bigl[ p_\mu p_\nu + g_{\mu\nu}(6m_e^2 - p^2)\bigr ]\,\frac{1}{\epsilon}  
\, , \nonumber \\
&
\overline{\G}_{A_\mu A_\nu}^{(1;1,0)}
  = \frac{-(4M^2 + M_A^2)\,p_\mu p_{\nu}
      + M^2\,\bigl[3(M^2 + M_A^2) + 4p^2\bigr] g_{\mu\nu}}
     {8\,\pi^2\,v^2}
\,\frac{1}{\epsilon}
  \, , \quad 
  \overline{\G}_{A_\mu A_\nu}^{(1;0,1)}
  =0 \, , \nonumber \\
&
\overline{\G}_{A_\mu A_\nu}^{(1;1,1)}
  = \frac{4\,(3M^2 + M_A^2)\,p_\mu p_{\nu}
      -\bigl[ 9\,(M^4 + M_A^2 M^2 + M_A^4)
                      +(12M^2 + M_A^2)\,p^2\bigr ] g_{\mu\nu}}
     {24\,\pi^2\,v^2}
\;\frac{1}{\epsilon}\, .
  \label{G.AA}
\end{alignat}

\subsubsection{The two-point fermion amplitude $\G^{(1)}_{\bar\psi \psi}$}

We introduce form factors as follows
\begin{align}
    \overline{\G}^{(1;\ln,\lV)}_{\bar\psi \psi(k)} = f^{(1;\ln,\lV)}_L(k^2) P_L + f^{(1;\ln,\lV)}_R(k^2) P_R + f^{(1;\ln,\lV)}_{LV}(k^2) P_L \slashed{k} + 
    f^{(1;\ln,\lV)}_{RV}(k^2) P_R \slashed{k} \, .
\end{align}
Then
\begin{alignat}{2}
f_L^{(1;0,0)}
  &= f_R^{(1;0,0)}
    = \frac{3\,m_e^3}{32\pi^2v^2}\,\frac{1}{\epsilon} \, ,
&\quad
f_{LV}^{(1;0,0)}(k^2)
  &= f_{RV}^{(1;0,0)}(k^2)
    = \frac{-k^2 + 7m_e^2}{64\pi^2v^2}\,\frac{1}{\epsilon} \, , \nonumber \\
f_L^{(1;0,1)}
  &= f_R^{(1;0,1)}
    = \Big[\frac{m_e^3}{32\pi^2v^2}
           -\frac{3\,e^2m_e}{32\pi^2}\Big]
      \,\frac{1}{\epsilon} \, ,
&\quad
f_{LV}^{(1;0,1)}(k^2)
  &= f_{RV}^{(1;0,1)}(k^2)
    = \frac{k^2 - 3m_e^2}{64\pi^2v^2}\,\frac{1}{\epsilon}, \nonumber \\
f_L^{(1;1,0)}
  &= f_R^{(1;1,0)}
    = -\,\frac{m_e^3}{8\pi^2v^2}\,\frac{1}{\epsilon} \, ,
&\quad
f_{LV}^{(1;1,0)}(k^2)
  &= f_{RV}^{(1;1,0)}(k^2)
    = \frac{m_e^2}{16\pi^2v^2}\,\frac{1}{\epsilon} \, , \nonumber \\
f_L^{(1;1,1)}
  &= f_R^{(1;1,1)}
    = 0 \, ,
&\quad
f_{LV}^{(1;1,1)}(k^2)
  &= f_{RV}^{(1;1,1)}(k^2)
    = 0 \, .
\label{G.psibarpsi}
\end{alignat}

\subsubsection{The three-point gauge-scalar amplitude $\G^{(1)}_{A_\mu A_\nu \sigma}$}

The UV divergent contribution $\overline{\G}^{(1;\ln,\lV)}_{A_\mu(p_1) A_\nu(p_2) \sigma}$ is written in terms of the following
form factors
\begin{align}
\overline{\G}^{(1;\ln,\lV)}_{A_\mu(p_1) A_\nu(p_2) \sigma} & = E^{(1;\ln,\lV)}_0 \, g_{\mu\nu} + E^{(1;\ln,\lV)}_1 \, p_{1\mu} p_{1\nu} + E^{(1;\ln,\lV)}_2 \, p_{2\mu} p_{2\nu} \nonumber \\
& \quad   + E^{(1;\ln,\lV)}_3 \, p_{1\mu} p_{2\nu} + E^{(1;\ln,\lV)}_4 \, p_{2\mu} p_{1\nu} \, .
\end{align}
In the sector $(0,0)$ the only non-vanishing form factor is
\begin{align}
    E_0^{(1;0,0)} = \frac{e^2 m_e^2}{2\pi^2\,v}\,g^{\mu\nu}\,\frac{1}{\epsilon} \, .
\end{align}
Moreover
\begin{align}
    E_0^{(1;1,0)} & = -\,\frac{p_1^2\bigl(M_A^2+8M^2\bigr)
      +2M_A^2\,(p_1\!\cdot\!p_2)
      +p_2^2 (M_A^2 + 8M^2) 
      +6M_A^2M^2
      +12M^4}
     {8\pi^2\,v^3}\;\frac{1}{\epsilon} \, , 
     \nonumber \\
    E_1^{(1;1,0)} & =
     \frac{-3\,p_1^2 + 2\,p_1\!\cdot\!p_2 + 6\,p_2^2 + 6\,M_A^2 + 33\,M^2}
     {24\,\pi^2\,v^3}\,\frac{1}{\epsilon} \, ,
     \nonumber \\
     E_2^{(1;1,0)} & = \frac{6\,p_{1}^{2} \;+\; 2\,p_{1}\!\cdot\!p_{2} \;-\; 3\,p_{2}^{2} \;+\; 6\,M_A^2 \;+\; 33\,M^{2}}
     {24\,\pi^{2}\,v^{3}} \,\frac{1}{\epsilon} \, , 
     \nonumber \\
     E_3^{(1;1,0)} & = \frac{-5\,p_1^2 \;-\; 12\,p_1\!\cdot\!p_2 \;-\; 5\,p_2^2 \;+\; 6\,M_A^2 \;+\; 18\,M^2}
     {24\,\pi^2\,v^3} \,\frac{1}{\epsilon} \, ,
      \nonumber \\
     E_4^{(1;1,0)} & = 0 \, , \nonumber \\
     E_0^{(1;2,0)} & = \frac{M^2\bigl(7\,p_1^2 - 10\,p_1\!\cdot\!p_2 + 7\,p_2^2 + 18\,M_A^2 + 27\,M^2\bigr)}
     {12\,\pi^2\,v^3}
      \;\frac{1}{\epsilon}\, , 
     \nonumber \\
     E_1^{(1;2,0)} & = E_2^{(1;2,0)} =
     -\,\frac{3\,M_A^2 + 20\,M^2}{24\,\pi^2\,v^3}\,\frac{1}{\epsilon}\, ,
     \qquad E_3^{(1;2,0)}  = E_4^{(1;2,0)} = \frac{M^2}{6\pi^2\,v^3}\,\frac{1}{\epsilon} \, ,
     \nonumber \\
    E_0^{(1;1,1)} & 
    = \frac{1}{24 \pi^2 v^3}\Big \{ p_1^4 + p_2^4 + 4 p_1^2 p_2^2 + 
    (18 M_A^2 + 40 M^2) (p_1^2 + p_2^2) \nonumber \\
    & \qquad \qquad \quad + 
    [ 3 (p_1^2 + p_2^2) + 28 M_A^2 - 16 M^2] p_1 \cdot p_2 
    + 72(M^4 + M_A^4 + M_A^2 M^2 )
    \Big \} \;\frac{1}{\epsilon}
\, ,
     \nonumber \\
    E_1^{(1;1,1)} & = \frac{5\,p_1^2 \;-\; 4\,p_1\!\cdot\!p_2 \;-\; 16\,p_2^2 \;-\; 45 M_A^2 \;-\; 61\,M^2}
     {24\,\pi^2\,v^3}
\;\frac{1}{\epsilon} \, ,
     \nonumber \\
     E_2^{(1;1,1)} & = \frac{-16\,p_1^2 \;-\; 4\,p_1\!\cdot\!p_2 \;+\; 5\,p_2^2 \;-\; 45 M_A^2 \;-\; 61\,M^2}
     {24\,\pi^2\,v^3} \, , 
     \nonumber \\
     E_3^{(1;1,1)} & = \frac{5\,p_1^2 + 14\,p_1\!\cdot\!p_2 + 5\,p_2^2 - 25\,M_A^2 - 23\,M^2}
     {12\,\pi^2\,v^3}
\,\frac{1}{\epsilon} \, ,
      \nonumber \\
     E_4^{(1;1,1)} & =  -\,\frac{3\,p_1^2 + 3\,p_2^2 + 8\,M_A^2 - 20\,M^2}
       {24\,\pi^2\,v^3}\;\frac{1}{\epsilon} \, , \nonumber \\
   E_0^{(1;2,1)} & 
    = -\,\frac{1}{24\,\pi^2\,v^3} \Big [ p_1^2\,(M_A^2 + 14M^2) \;+\; M_A^2\,p_2^2 \;-\; 20M^2\,p_1\!\cdot\!p_2 \nonumber \\
    & \qquad \qquad \quad
      \;+\; 14M^2\,p_2^2 \;+\; 36 M_A^2 M^2 \;+\; 18 M_A^4 \;+\; 54M^4 \Big ]   
\,\frac{1}{\epsilon} \, ,
     \nonumber \\
    E_1^{(1;2,1)} & =  E_2^{(1;2,1)} = 
    \frac{M_A^2 + 5\,M^2}{6\pi^2\,v^3}\,\frac{1}{\epsilon}
\, , \qquad 
     E_3^{(1;2,1)}  = E_4^{(1;2,1)} = -\,\frac{M^2}{6\pi^2\,v^3}\,\frac{1}{\epsilon} \, ,
      \nonumber \\
    E_0^{(1;1,2)} & 
    = -\,\frac{1}{24\,\pi^2\,v^3} \Big [ 
  p_1\!\cdot\!p_2\bigl(3p_2^2 + 22e^2v^2 - 16M^2\bigr)
  + p_1^2\bigl(3p_1\!\cdot\!p_2 + 4p_2^2 + 14 M_A^2 + 16M^2\bigr)
  \nonumber \\
  & \qquad \qquad
  + 14 M_A^2\,p_2^2 + 16M^2\,p_2^2 + p_1^4 + p_2^4
  + 54M_A^2 M^2 + 72 M_A^4 + 36M^4 \Big ] 
\;\frac{1}{\epsilon}
\, ,
     \nonumber \\
    E_1^{(1;1,2)} & =  \frac{-p_1^2 + p_1\!\cdot\!p_2 + 5\,p_2^2 + 19\,M_A^2 + 14\,M^2}
     {12\,\pi^2\,v^3}
\;\frac{1}{\epsilon} \, ,
     \nonumber \\
     E_2^{(1;1,2)} & = \frac{5\,p_1^2 + p_1\!\cdot\!p_2 - p_2^2 + 19\,M_A^2 + 14\,M^2}
     {12\,\pi^2\,v^3}
\,\frac{1}{\epsilon} \, , 
     \nonumber \\
     E_3^{(1;1,2)} & = \frac{-5\,p_1^2 \;-\;16\,p_1\!\cdot\!p_2 \;-\;5\,p_2^2 \;+\;44\,M_A^2 \;+\;28\,M^2}
     {24\,\pi^2\,v^3}
\,\frac{1}{\epsilon} \, ,
      \nonumber \\
     E_4^{(1;1,2)} & =  \frac{3\,p_1^2 + 3\,p_2^2 + 8\,M_A^2 - 20\,M^2}
     {24\,\pi^2\,v^3}
\,\frac{1}{\epsilon} \, . \nonumber \\
   \label{G.sigmaAA}
\end{align}
The form factors in the sectors $(0,1)$, $(0,2)$, $(3,0)$, $(0,3)$ vanish.

\subsubsection{The three-point gauge-Goldstone-scalar amplitude $\G^{(1)}_{A_\mu \chi \sigma}$}

The UV divergent contribution $\overline{\G}^{(1;\ln,\lV)}_{A_\mu \chi \sigma}$ is written in terms of the following form factors
\begin{align}
    \overline{\G}^{(1;\ln,\lV)}_{A_\mu(-p_1) \chi(-p_2) \sigma} = 
    H_0^{(1;\ln,\lV)} p_1^\mu +  H_1^{(1;\ln,\lV)} p_2^\mu \, .
    \label{G.sigmaAchi}
\end{align}
We have
\begin{align}
    & H_0^{(1;0,0)} = -\,\frac{i\,e\,m_e^2}{4\pi^2\,v^2}\,\frac{1}{\epsilon} \, , 
    \qquad
    H_1^{(1;0,0)} = -\,\frac{i\,e\,m_e^2}{2\pi^2\,v^2}\,\frac{1}{\epsilon} \, , 
    \nonumber \\
    & H_0^{(1;1,0)} = -\,\frac{i}{24\,\pi^2\,e\,v^4}
\;\bigl[
  (p_1\!\cdot\!p_2)\,\bigl(-6\,p_2^2 + 6\,M_A^2 + 33\,M^2\bigr)
  + 6\,M_A^2\,p_2^2
  - p_1^2\,\bigl(3\,p_1\!\cdot\!p_2 + 5\,p_2^2 + 3\,M_A^2\bigr) \nonumber \\
  & \qquad\qquad\qquad\qquad\qquad
  + 18\,M^2\,p_2^2
  + 2\,(p_1\!\cdot\!p_2)^2
  - 5\,p_2^4
  + 9\,M_A^2\,M^2
  + 9\,M^4
\bigr]
\;\frac{1}{\epsilon} \, , \nonumber \\
    & H_1^{(1;1,0)} = \frac{i}{24\,\pi^2\,e\, v^4}
\Bigl[
3p_1^2\bigl(-2p_2^2 + M_A^2 + 8M^2\bigr)
+3\bigl(-p_2^2(M_A^2+3M^2) + 3 p_2^4 \nonumber \\
& \qquad\qquad\qquad\qquad\qquad+ 6M^2(M_A^2+2M^2)\bigr)
+p_1\!\cdot\!p_2\,(6M_A^2-2p_2^2)
\Bigr]
\,\frac{1}{\epsilon} \, , \nonumber \\
    & H_0^{(1;2,0)} = i \frac{p_1\!\cdot\!p_2\,(3M_A^2 + 20M^2) - 4M^2\,p_2^2}
     {24\,\pi^2\,e\,v^4}\,\frac{1}{\epsilon} \, , \nonumber \\
    & H_1^{(1;2,0)} =  -\,\frac{i}{24\,e\,v^4\, \pi^2}
\bigl[
-3\,M_A^2\,p_2^2 + 14\,M^2\,p_1^2 - 16\,M^2\,p_1\!\cdot\!p_2 - 6\,M^2\,p_2^2\nonumber \\
& \qquad\qquad\qquad\qquad\quad  + 36\,M_A^2\,M^2\, + 54\,M^4
\bigr]
\,\frac{1}{\epsilon} \, , 
\nonumber \\
    & H_0^{(1;1,1)} =  \frac{i}{24\,e\,\pi^2\,v^4}
\Bigl[
  (p_1\!\cdot\!p_2)\,(-12\,p_2^2 + 45\,M_A^2 + 61\,M^2)
  + 50\,M_A^2\,p_2^2 \nonumber \\
  & \qquad\qquad\qquad\qquad\quad
  - p_1^2\,(5\,p_1\!\cdot\!p_2 + 10\,p_2^2 + 3\,M_A^2)
  + 46\,M^2\,p_2^2
  + 4\,(p_1\!\cdot\!p_2)^2 \nonumber \\
  & \qquad\qquad\qquad\qquad\quad  
  - 10\,p_2^4
  + 9\,M_A^2\,M^2
  + 9\,M_A^4 
  + 9\,M^4
\Bigr]
\,\frac{1}{\epsilon} \, , \nonumber \\
    & H_1^{(1;1,1)} =- \,\frac{i}{24\,e\,v^4\,\pi^2}
\Bigl[
2\,p_1^2\bigl(-6\,p_2^2 + 9\,M_A^2 + 20\,M^2\bigr)
+4\,(p_1\!\cdot\!p_2)\bigl(-p_2^2 + 5\,M_A^2 + M^2\bigr) \nonumber \\
  & \qquad\qquad\qquad\qquad\quad
-3\,p_2^2\bigl(9\,M_A^2 + 7\,M^2\bigr)
+ p_1^4
+ 6\,p_2^4
+ 72\,(M_A^2\,M^2\, + M_A^4 + M^4)
\Bigr]
\,\frac{1}{\epsilon} \, , \nonumber \\
    & H_0^{(1;2,1)} =
-\,i \frac{p_1\!\cdot\!p_2\,(M_A^2 + 5M^2) - M^2\,p_2^2}
      {6\,\pi^2\,e\,v^4}\;\frac{1}{\epsilon}
\, , \nonumber \\
    & H_1^{(1;2,1)} = \frac{i}{24\,e\,v^4\pi^2}
\Bigl[
  p_1^2\,(M_A^2 + 14M^2)
  \;-\;3\,M_A^2\,p_2^2
  \;-\;16\,M^2\,(p_1\!\cdot\!p_2)
  \;-\;6\,M^2\,p_2^2 \nonumber \\
  & \qquad\qquad\qquad\qquad\quad  
  \;+\;36\,M_A^2 M^2
  \;+\;18\,M_A^4
  \;+\;54\,M^4
\Bigr]
\,\frac{1}{\epsilon} \, , \nonumber \\
    & H_0^{(1;1,2)} = -\,\frac{i}{24\,e\,v^4\, \pi^2}
\Bigl[
  (p_1\!\cdot\!p_2)\bigl(-2\,p_1^2 - 6\,p_2^2 + 38\,M_A^2 + 28\,M^2\bigr)
  \nonumber \\
  & \qquad\qquad\qquad\qquad\quad   
  + p_2^2\bigl(-5\,p_1^2 - 5\,p_2^2 + 44\,M_A^2 + 28\,M^2\bigr)
  + 2\,(p_1\!\cdot\!p_2)^2
\Bigr]
\,\frac{1}{\epsilon} \, , \nonumber \\
& H_1^{(1;1,2)} = \frac{i}{24\,e\,v^4\,\pi^2}
\Bigl\{
  3\bigl[ -4\,p_2^2\,(2 M_A^2 + M^2) + p_2^4 + 6\,(3 M_A^2 M^2 + 4 M_A^4 + 2M^4)\bigr ]
  \nonumber \\
  &\qquad\qquad\qquad\qquad\quad
  +2\,p_1^2\bigl(-3p_2^2 + 7M_A^2 + 8M^2\bigr)
  +2\,p_1\!\cdot\!p_2 \bigl(-p_2^2 + 7M_A^2 + 2M^2\bigr)
  +p_1^4
\Bigr\}\frac{1}{\epsilon} \, .
\label{G.Achisigma}
\end{align}
The form factors in the sectors $(0,1)$, $(0,2)$, $(3,0)$, $(0,3)$ vanish.

\subsubsection{The three point scalar-Goldstone amplitude $\G^{(1)}_{\sigma(q) \chi(p_1) \chi(p_2)}$}
The sector decomposition is given by
($q=-p_1-p_2$)
\begin{align}
& \overline{\G}^{(1;0,0)}_{\sigma(q) \chi(p_1) \chi(p_2)} = 
-\frac{16 m_e^4 + 3 q^4  }{16 \pi^2 v^3} \frac{1}{\epsilon} \, , \nonumber \\
& \overline{\G}^{(1;1,0)}_{\sigma(q) \chi(p_1) \chi(p_2)} = 
\Big [ 
- \frac{M^4}{2 \pi^2 v^3}
-\frac{M^2 ( 3 M^2 + 4 M_A^2)}{8 \pi^2}\frac{p_1^2 + p_2^2}{e^2 v^5}
+ \frac{M^2( 3 M^2 + M_A^2)}{2 \pi^2}
\frac{p_1 \cdot p_2}{ e^2 v^5}
\nonumber \\
& \qquad \qquad \qquad \qquad
- \frac{3 M^2 + M_A^2}{8 \pi^2} \frac{(p_1^2 + p_2^2) ~p_1 \cdot p_2 }{e^2 v^5}
- \frac{(3 M^2 + M_A^2)}{4 \pi^2}\frac{p_1^2 p_2^2}{e^2 v^5}
+ \frac{(p_1 \cdot p_2)^2}{4 \pi^2 v^3} 
+ \frac{p_1^4 + p_2^4}{8 \pi^2 v^3} 
\nonumber \\
& \qquad \qquad \qquad \qquad 
+ 
\frac{5}{24\pi^2}
\frac{p_1^2 p_2^2 ~(p_1^2+p_2^3)}{e^2 v^5 }
+\frac{1}{8 \pi^2}\frac{p_1\cdot p_2~ (p_1^4 + p_2^4)}{e^2 v^5}
-\frac{1}{12 \pi^2}\frac{(p_1 \cdot p_2)^2 ~(p_1^2 + p_2^2)}{e^2 v^5}
\Big ]
\frac{1}{\epsilon}
\, , \nonumber \\
& \overline{\G}^{(1;0,1)}_{\sigma(q) \chi(p_1) \chi(p_2)} =  
-\frac{24 M_A^2 + 9 M_A^2 q^2 - 2 q^4}{8 \pi^2 v^3}  \frac{1}{\epsilon} \, , \nonumber \\
& \overline{\G}^{(1;2,0)}_{\sigma(q) \chi(p_1) \chi(p_2)} =  
 \Big [
\frac{3 M^4}{4\pi^2 v^3} 
+ \frac{M^2}{8 \pi^2 v^3}(p_1^2 + p_2^2)
-  \left( \frac{9 M^4}{4 \pi^2 e^2 v^5} + \frac{5 M^2}{4 \pi^2 v^3} \right)
p_1 \cdot p_2 \nonumber \\
& \qquad \qquad \qquad \qquad
+ \left ( \frac{M^2}{4 \pi^2 e^2 v^5} + \frac{1}{8 \pi^2 v^3} \right )
( p_1^2 + p_2^2) (p_1 \cdot p_2) 
+ \frac{2 M^2}{3 \pi^2 e^2 v^5} (p_1 \cdot p_2)^2
- \frac{M^2}{6\pi^2 e^2 v^5} p_1^2 p_2^2
\Big ] \frac{1}{\epsilon} \, , \nonumber \\
& \overline{\G}^{(1;1,1)}_{\sigma(q) \chi(p_1) \chi(p_2)} =
\Big [
\frac{3( M^4 + M^2 M_A^2 + M_A^4)}{8 \pi^2}\frac{p_1^2 + p_2^2}{e^2 v^5}
- \frac{3(M^4 + M^2 M_A^2 + M_A^4  )}{\pi^2}
\frac{p_1 \cdot p_2}{ e^2 v^5}
\nonumber \\
& \qquad \qquad \qquad \qquad
+ \frac{(7 M^2 + 9 M_A^2)}{8 \pi^2} \frac{(p_1^2 + p_2^2) ~p_1 \cdot p_2 }{e^2 v^5}
+ \frac{(23 M^2 + 25 M_A^2)}{12 \pi^2}\frac{p_1^2 p_2^2}{e^2 v^5}
- \frac{(M^2+5 M_A^2)}{e^2 v^5}  \frac{(p_1 \cdot p_2)^2}{6 \pi^2}
\nonumber \\
& \qquad \qquad \qquad \qquad 
- \frac{p_1^4 + p_2^4}{8 \pi^2 v^3} -
\frac{5}{12\pi^2}
\frac{p_1^2 p_2^2 ~(p_1^2+p_2^3)}{e^2 v^5 }
-\frac{1}{4\pi^2}\frac{p_1\cdot p_2~ (p_1^4 + p_2^4)}{e^2 v^5}
+\frac{1}{6\pi^2}\frac{(p_1 \cdot p_2)^2 ~(p_1^2 + p_2^2)}{e^2 v^5}
\Big ]
\frac{1}{\epsilon}
%
\, , \nonumber \\
& \overline{\G}^{(1;0,2)}_{\sigma(q) \chi(p_1) \chi(p_2)} = 
 \frac{1}{16 \pi^2 v^3} \left( 60 M_A^4 + 12 M_A^2 q^2 - q^4 \right)
 \frac{1}{\epsilon}\, , \nonumber \\
& \overline{\G}^{(1;3,0)}_{\sigma(q) \chi(p_1) \chi(p_2)} = 0 \, , \nonumber \\
& \overline{\G}^{(1;2,1)}_{\sigma(q) \chi(p_1) \chi(p_2)} =  
\Big [ 
\frac{3}{4 \pi^2} \frac{(3 M^4 + 2 M^2 M_A^2 + M_A^4)}{e^2 v^5}
p_1 \cdot p_2
\nonumber \\
& \qquad \qquad \qquad \qquad
- \frac{2 M^2 + M_A^2}{8 \pi^2} \frac{(p_1^2 + p_2^2) ~p_1 \cdot p_2 }{e^2 v^5}
+ \frac{ M^2 }{6 \pi^2}\frac{p_1^2 p_2^2}{e^2 v^5}
- \frac{2M^2}{3\pi^2}\frac{(p_1 \cdot p_2)^2}{ e^2 v^5} 
\Big ]
\frac{1}{\epsilon}
\, , \nonumber \\
& \overline{\G}^{(1;1,2)}_{\sigma(p) \chi(p_1) \chi(p_2)} = 
\Big [
 \frac{3(2 M^4 + 3 M^2 M_A^2 +4 M_A^4  )}{4\pi^2}
\frac{p_1 \cdot p_2}{ e^2 v^5}
\nonumber \\
& \qquad \qquad \qquad \qquad
- \frac{(M^2 + 2 M_A^2)}{2 \pi^2} \frac{(p_1^2 + p_2^2) ~p_1 \cdot p_2 }{e^2 v^5}
- \frac{(7 M^2 + 11 M_A^2)}{6 \pi^2}\frac{p_1^2 p_2^2}{e^2 v^5}
+ \frac{(2 M^2+7 M_A^2)}{e^2 v^5}  \frac{(p_1 \cdot p_2)^2}{12 \pi^2}
\nonumber \\
& \qquad \qquad \qquad \qquad 
 +
\frac{5}{24\pi^2}
\frac{p_1^2 p_2^2 ~(p_1^2+p_2^3)}{e^2 v^5 }
+\frac{1}{8\pi^2}\frac{p_1\cdot p_2~ (p_1^4 + p_2^4)}{e^2 v^5}
-\frac{1}{12\pi^2}\frac{(p_1 \cdot p_2)^2 ~(p_1^2 + p_2^2)}{e^2 v^5}
\Big ]
\frac{1}{\epsilon}
\, , \nonumber \\
& \overline{\G}^{(1;0,3)}_{\sigma(p) \chi(p_1) \chi(p_2)} = 0 \,  .
\end{align}

\subsubsection{The three-point fermion-Goldstone amplitude $\G^{(1)}_{\chi(q) \bar\psi(p_1) \psi(p_2)}$}

We introduce the form factor decomposition

\begin{align}
    \overline{\G}^{(1;\ln,\lV)}_{\chi(q) \bar\psi(p_1) \psi(p_2)} = & \quad F^{(1;\ln,\lV)}_{L}(q,p_1,p_2) P_L +
                                   F^{(1;\ln,\lV)}_{R}(q,p_1,p_2) P_R \nonumber \\
                                   & + F^{(1;\ln,\lV)}_{LV}(q,p_1,p_2) P_L \slashed{q} +
                                   F^{(1;\ln,\lV)}_{RV}(q,p_1,p_2) P_R \slashed{q} + F^{(1;\ln,\lV)}_{T;\alpha \beta}(q,p_1,p_2) \sigma^{\alpha\beta} \, .
\end{align}

The form factors are
\begin{alignat}{2}
F_L^{(1;0,0)}
  &= - F_R^{(1;0,0)} = 
    \frac{i ~ m_e}{64 \pi^2  v^3}
  \Big (
  q^2 + 6 m_e^2
  \Big ) \,\frac{1}{\epsilon} \, ,
&\quad
F_{LV}^{(1;0,0)}
  &= F_{RV}^{(1;0,0)}
    = -\frac{3 i \, m_e^2}{64 \pi^2 v^3 } \,\frac{1}{\epsilon} \, , \nonumber \\
F_{T;\alpha\beta}^{(1;0,0)} &= -\frac{i \, m_e}{64 \pi^2 v^3}
\epsilon^{\alpha\beta\mu\nu}p_{1\mu}p_{2\nu} \,\frac{1}{\epsilon} \, , & \nonumber \\
F_L^{(1;0,1)}
  &= - F_R^{(1;0,1)} = -
    \frac{i ~ m_e}{64 \pi^2  v^3}
  \Big (
  q^2 + 6 M_A^2 -2  m_e^2
  \Big ) \,\frac{1}{\epsilon} \, ,
&\quad
F_{LV}^{(1;0,1)}
  &= F_{RV}^{(1;0,1)}
    = -\frac{i \, m_e^2}{64 \pi^2 v^3 } \,\frac{1}{\epsilon} \, , \nonumber \\
F_{T;\alpha\beta}^{(1;0,1)} &= \frac{i \, m_e}{64 \pi^2 v^3}
\epsilon^{\alpha\beta\mu\nu}p_{1\mu}p_{2\nu} \,\frac{1}{\epsilon} \, , & \nonumber \\
F_L^{(1;1,0)}
  &= - F_R^{(1;1,0)} = -
    \frac{i ~  m_e}{32 \pi^2  v^3}
(
  3 q^2  + 4 m_e^2
) \,\frac{1}{\epsilon} \, ,
&\quad
F_{LV}^{(1;1,0)}
  &= F_{RV}^{(1;1,0)}
    = \frac{i \, m_e^2}{4 \pi^2 v^3 } \,\frac{1}{\epsilon} \, , \nonumber \\
F_{T;\alpha\beta}^{(1;1,0)} &= -\frac{i \, m_e}{8\pi^2 v^3}
\epsilon^{\alpha\beta\mu\nu}p_{1\mu}p_{2\nu} \,\frac{1}{\epsilon} \, , & \nonumber \\
F_L^{(1;1,1)}
  &= - F_R^{(1;1,1)} =
  \frac{3 i ~ m_e q^2}{32 \pi^2  v^3}\,\frac{1}{\epsilon} \, ,
&\quad
F_{LV}^{(1;1,1)}
  &= F_{RV}^{(1;1,1)}
    = -\frac{3 i \, m_e^2}{16 \pi^2 v^3 } \,\frac{1}{\epsilon} \, , \nonumber \\
F_{T;\alpha\beta}^{(1;1,1)} &= \frac{i \, m_e}{8 \pi^2 v^3}
\epsilon^{\alpha\beta\mu\nu}p_{1\mu}p_{2\nu} \,\frac{1}{\epsilon} \, . &
    \nonumber \\
\label{G.chipsibarpsi}
\end{alignat}

\subsubsection{The three-point fermion-gauge amplitude $\G^{(1)}_{A_\mu(q) \bar\psi(p_1) \psi(p_2)}$}

In a similar way we define ($q=-p_1-p_2$)
\begin{align}
    \overline{\G}^{(1;\ln,\lV)}_{A_\mu(q) \bar\psi(p_1) \psi(p_2)} = & \quad G_{L}^{(1;\ln,\lV)}P_L q^\mu + G_{R}^{(1;\ln,\lV)}P_R q^\mu 
                                   + G_{LV}^{(1;\ln,\lV)}P_L \gamma^\mu + G_{RV}^{(1;\ln,\lV)} P_R \gamma^\mu \nonumber \\
                                   & + G_T^{(1;\ln,\lV)}\, \sigma^{\alpha\beta}\epsilon_{\mu\alpha\beta\nu}(p_1 - p_2)^\nu + i \widetilde{G}^{(1;\ln,\lV)} \, \epsilon^{\mu\nu\alpha\beta}\gamma_\nu p_{1\alpha} p_{2\beta} \, .
\end{align}

The form factors are 
\begin{alignat}{2}
G_L^{(1;0,0)} &= -G_R^{(1;0,0)} = \frac{e m_e}{64 \pi^2 v^2} \, , \qquad
G_T^{(1;0,0)}  = -\frac{e m_e}{128 \pi^2 v^2} \,,  \quad \widetilde{G}^{(1;0,0)} = \frac{e}{128 \pi^2 v^2} \, ,  \nonumber \\
G_{LV}^{(1;0,0)} &= -G_{RV}^{(1;0,0)}=\frac{e}{128 \pi^2 v^2} (p_1^2 + p_2^2 + p_1 \cdot p_2 - m_e^2 ) \, , \nonumber \\
G_L^{(1;0,1)} &= -G_R^{(1;0,1)} = -\frac{3\, e m_e}{32 \pi^2 v^2} \, , \qquad
G_T^{(1;0,1)}  = \frac{e m_e}{64 \pi^2 v^2} \, ,\quad \widetilde{G}^{(1;0,1)} = 0  \, , \nonumber \\
G_{LV}^{(1;0,1)} &= -G_{RV}^{(1;0,1)}=-\frac{9 \,e m_e^2}{32\pi^2 v^2} \, , \nonumber \\
G_L^{(1;1,0)} &= -G_R^{(1;1,0)} = -\frac{e m_e}{64 \pi^2 v^2} \, , \qquad
G_T^{(1;1,0)}  = \frac{e m_e}{128 \pi^2 v^2} \, ,\quad \widetilde{G}^{(1;1,0)} = -\frac{e}{128\pi^2v^2}  \, , \nonumber \\
G_{LV}^{(1;1,0)} &= -G_{RV}^{(1;1,0)}=-\frac{e}{128 \pi^2 v^2} (p_1^2 + p_2^2 + p_1 \cdot p_2 - m_e^2 ) \, ,
\nonumber \\
G_L^{(1;1,1)} &= -G_R^{(1;1,1)} = \frac{3 ~ e m_e}{32 \pi^2 v^2} \, , \qquad
G_T^{(1;1,1)}  = -\frac{e m_e}{64 \pi^2 v^2} \, ,\quad \widetilde{G}^{(1;1,1)} = 0 \, , \nonumber \\
G_{LV}^{(1;1,1)} &= -G_{RV}^{(1;1,1)}=\frac{3 \, e m_e^2}{16 \pi^2 v^2} \, .
\nonumber \\
\label{G.Atpsibarpsi}
\end{alignat}

\bibliography{bibliography_1loop}

\end{document}